\documentclass[prd,nofootinbib,floats,superscriptaddress,eqsecnum,tightenlines,preprintnumbers,11pt]{revtex4-1}

\usepackage{mathtools}
\usepackage{amssymb}
\usepackage{dsfont}
\usepackage{lipsum}
\usepackage{mathtools}
\usepackage{xcolor}
\usepackage[colorlinks=true,linkcolor=blue,citecolor=magenta,linktocpage=true]{hyperref}
\usepackage{physics}
\usepackage{csquotes}

\usepackage{caption}
\usepackage{subcaption}

\usepackage[makeroom]{cancel}

\def\be{\begin{equation}}
	\def\ee{\end{equation}}
\def\bea{\begin{eqnarray}}
	\def\eea{\end{eqnarray}}
\def\beq{\begin{eqnarray}}
	\def\eeq{\end{eqnarray}}
\def\bas{\begin{subequations}\begin{eqnarray}}
		\def\eas{\end{eqnarray}\end{subequations}}
\def\nn{\nonumber}

\newcommand{\cB}{{\mathcal B}}

\newcommand{\cI}{{\mathcal I}}

\newcommand{\cK}{{\mathcal K}}
\newcommand{\cL}{{\mathcal L}}

\newcommand{\cR}{{\mathcal R}}
\newcommand{\cO}{{\mathcal O}}

\newcommand{\cT}{{\mathcal T}}

\def\rd{\textrm{d}}

\newcommand{\bes}{\begin{eqnarray}}
	\newcommand{\ees}{\end{eqnarray}}

\def\ic{\vartheta}

\def\dd{\mathrm{d}}

\def\nn{\nonumber}

\begin{document}

\title{Nonlinear gravitational waves in Horndeski gravity: \\ \smallskip Scalar pulse and memories}

	\author{Jibril Ben Achour}
	\affiliation{Arnold Sommerfeld Center for Theoretical Physics, Munich, Germany}
	\affiliation{Munich Center for Quantum Science and Technology, Munich, Germany}
	\affiliation{Univ de Lyon, ENS de Lyon, Laboratoire de Physique, CNRS UMR 5672, Lyon 69007, France}
	\author{Mohammad Ali Gorji}
	\affiliation{Departament de F\'{i}sica Qu\`{a}ntica i Astrof\'{i}sica, Facultat de F\'{i}sica, Universitat de Barcelona,
		Mart\'{i} i Franqu\`{e}s 1, 08028 Barcelona, Spain}
	\author{Hugo Roussille}
	\affiliation{Univ de Lyon, ENS de Lyon, Laboratoire de Physique, CNRS UMR 5672, Lyon 69007, France}

%\date{\today}

	\begin{abstract}
		We present and analyze a new \textit{non-perturbative radiative} solution of Horndeski gravity. This exact solution is constructed by a disformal mapping of a seed solution of the shift-symmetric Einstein-Scalar system belonging to the Robinson-Trautman geometry describing the gravitational radiation emitted by a time-dependent scalar monopole. After analyzing in detail the properties of the seed, we show that while the general relativity solution allows for shear-free parallel transported null frames, the disformed solution can only admit parallel transported null frames with a non-vanishing shear.
		This result shows that, at the nonlinear level, the scalar-tensor mixing descending from the higher-order terms in Horndeski dynamics can generate shear out of a pure scalar monopole. We further confirm this analysis by identifying the spin-0 and spin-2 polarizations in the disformed solution using the Penrose limit of our radiative solution. 
		Finally, we compute the geodesic motion and the memory effects experienced by two null test particles with vanishing initial relative velocity after the passage of the pulse. 
		This exact radiative solution offers a simple framework to witness nonlinear consequences of the scalar-tensor mixing in higher-order scalar-tensor theories.
	\end{abstract}

\maketitle
\tableofcontents

\newpage

\section{Introduction}
	
	With the advent of gravitational wave (GW) astronomy, characterizing the mechanisms of production and propagation of GWs in modified theories of gravity stands as a crucial task in order to probe possible deviations w.r.t. to general relativity (GR). Scalar-tensor theories are by far the most studied alternatives candidates so far. The presence of the additional scalar field manifests itself through a new scalar polarization for the GW which can mix in a complicated way with the tensorial sector, leading to a rich phenomenology \cite{ Kobayashi:2011nu, Saltas:2014dha, Nishizawa:2017nef, Hou:2017bqj, Nunes:2018zot, Creminelli:2018xsv, Garoffolo:2019mna, Dalang:2019rke, Dalang:2020eaj, Kubota:2022lbn, Battista:2021rlh}. At the cosmological level, the leading effects of these scalar-tensor mixing on GWs propagating over cosmological distances result in a damped amplitude and a change of phase velocity of the wave. Confronting this phenomenology with current data allows one to drastically constrain alternative theories of gravity operating at this energy scale and design suitable probes for dark energy models \cite{Nishizawa:2019rra, Arai:2017hxj, Kase:2018aps, Quartin:2023tpl, Langlois:2017dyl, Gong:2017kim, Takeda:2021hgo, Takeda:2023wqn}. For compact objects, the scalar-tensor mixing also manifests in many ways, modifying the waveforms at the different stages of the coalescence of a binary system, from the merger phase to the quasi-normal modes spectrum up to the memory contribution \cite{Lang:2013fna, Lang:2014osa, Bernard:2018hta, Bernard:2018ivi, AbhishekChowdhuri:2022ora, Bernard:2022noq, Higashino:2022izi,Du:2016hww,  Koyama:2020vfc, Tahura:2020vsa, Hou:2020wbo, Hou:2020tnd, Hou:2020xme, Seraj:2021qja, Gorji:2022hyy, Heisenberg:2023prj}. Moreover, the presence of a scalar charge for compact objects also affects the orbital motion and the waveform emitted during extreme-mass ratio inspirals, as recently discussed in \cite{Maselli:2020zgv, Barsanti:2022ana}. Therefore, identifying the consequences of these scalar-tensor mixing and the resulting phenomenology provides a powerful window to constrain new physics beyond GR.
	
	To a large extent, the phenomenology of GWs in scalar-tensor theories has been investigated at the perturbative level. Yet, a detailed characterization of the GWs in these scalar-tensors theories requires a non-perturbative analysis. This can be addressed in two ways. A first approach consists in investigating how the asymptotic BMS symmetries realized at $\cI^{+}$ are modified w.r.t. GR \cite{Flanagan:2015pxa, Compere:2019gft, Freidel:2021fxf}. In particular, the possibility for matter to couple to a conformally or disformally related metric can change the falloff behavior of the scalar field profile near $\cI^{+}$, while the new interaction terms beyond GR will modify the expressions of both the charges and fluxes appearing in the flux-balance laws. This approach was investigated for the Brans-Dicke theory in \cite{Tahura:2020vsa, Hou:2020wbo, Hou:2020tnd, Hou:2020xme, Seraj:2021qja}. To our knowledge, no higher-order modified gravity theories was investigated along this line so far.  A second road, less ambitious, consists in constructing exact radiative solution of modified gravity theories and investigate the new phenomenology showing up in this nonlinear regime. In this work, we follow the second approach and present a new exact radiative solution in the Horndeski theory, revealing in that way a new surprising phenomenology.
	
	Already in GR, exact radiative solutions have been known for a long time, such as pp-waves and their gyratonic extension, colliding plane wave solutions, as well as Kundt and Robinson-Trautman geometries. While they are not relevant to confront theory to observations, they nevertheless serve as theoretical laboratories to explore the exact nonlinear features of the radiative regime. The nonlinear memory effects stand as one key example which can be investigated analytically in these exact solutions \cite{Zhang:2017rno, Zhang:2018srn, Chakraborty:2022qvv, Chakraborty:2019yxn, Shore:2018kmt,Flanagan:2019ezo}. Another interesting example stems from the properties of exact solutions describing colliding gravitational plane waves. These solutions of Einstein equations allow for the formation of trapped regions and singularities, providing a framework to investigate the formation of black holes from interacting (idealized) GWs at the fully nonlinear level \cite{Yoshino:2007ph, Pretorius:2018lfb}. They also enjoy interesting dualities with exact black hole solutions.  Moreover, they offer a framework to investigate the effects of the gravitational field of ultra-relativistic particles in scattering processes, providing elegant relations between gravitational observables and scattering amplitudes in the high energy regime \cite{tHooft:1987vrq, Arefeva:1994pzt}. Finally, let us mention the rich literature on the exact solitonic solutions in GR \cite{Belinski:2001ph, Manzoni:2021dij}. At the cosmological level, these inhomogeneous cosmological solitonic wave solutions provide interesting models for a non-perturbative treatment of early universe cosmology \cite{Belinsky:1979wi, Carr:1983jzn, Verdaguer:1986bq}, with interesting applications to dark energy \cite{Belinski:2017luc}.  In view of this rich phenomenology, it is an interesting challenge to also explore the fully nonlinear radiative regime of modified gravity theories by constructing exact radiative solutions.
	
	%The most direct path to explore this regime consists in constructing exact radiative solutions in scalar-tensor theories. 
	For simple enough theories such as Brans-Dicke gravity, exact radiative solutions have been worked out \cite{Siddhant:2020gkn, Nozawa:2023boa}. However, when considering higher-order scalar-tensor theories, constructing exact radiative solutions turns out to be a difficult task due to the high complexity of the field equations (see \cite{Babichev:2012qs, Ayon-Beato:2005gdo, Kolar:2021uiu, Zhang:2021sjx} for some examples). One way to bypass this difficulty relies on using solution-generating techniques. For degenerate higher-order scalar-tensor (DHOST) theories \cite{Langlois:2015cwa, BenAchour:2016fzp}, disformal field redefinition %introduced by Bekenstein in \cite{Bekenstein:1992pj} 
	can be used to explore both the theory and solution spaces in an efficient way \cite{Zumalacarregui:2013pma, Bettoni:2013diz, BenAchour:2016cay}. Provided a given seed solution is known for a given DHOST theory, one can generate a new exact solution for another DHOST theory belonging to the same degeneracy class (which are stable under disformal field redefinition). This trick was first explored in \cite{BenAchour:2020wiw} and further used in \cite{Faraoni:2021gdl, Bakopoulos:2022csr, BenAchour:2020fgy, Anson:2020trg, Baake:2021kyg} to construct both non-stealth spherically symmetric and rotating black hole solutions (see also \cite{Minamitsuji:2020jvf, Minamitsuji:2021rtw, Domenech:2015hka, Tsujikawa:2015upa, Domenech:2015tca, Fujita:2015ymn} for related works on stars and cosmological perturbations). While being a pure field redefinition, this map generates new physics when one assumes that matter minimally couples to the disformed metric. When analyzing the kinematical properties of a given geometry, such as its geodesics, its causal structure through its null vectors, one implicitly assumes the existence of an observer which will attribute new properties to the disformed geometry. In particular, the algebraic characterization of the disformed geometry based on the identification of principal null directions (PNDs) which determines the Petrov type of the geometry can dramatically change under a disformal transformation, as shown in \cite{Achour:2021pla}. Thus, this solution generating map allows one to reveal in a simple way new phenomenology appearing in modified gravity. In this work, we shall use this disformal solution-generating map to construct a first exact radiative solution (beyond pp-wave) for a specific Horndeski theory and analyze the nonlinear effects of the scalar-tensor mixings on the GWs.
	
	As a starting point, we will consider the Einstein-Scalar system as a seed theory. For this system, an exact radiative solution has been derived in \cite{Tahamtan:2015sra} and further analyzed in \cite{Tahamtan:2016fur}. This solution will serve as the seed solution in our construction. As we shall see, this geometry describes a time-dependent scalar monopole which takes the form of a pulse. The kinetic energy of the scalar field grows up to a maximum before decaying, radiating in the process \enquote{spherical} GWs, i.e. purely longitudinal waves. This exact GR solution belongs to the Robinson-Trautman family of solutions of Petrov type II. The first part of this work therefore consists in a detailed analysis of this exact solution of GR, completing and improving the analysis presented in \cite{Tahamtan:2015sra, Tahamtan:2016fur}. Let us stress that the key interest in this solution is that it allows one to isolate the source of radiation to a pure scalar monopole. When the scalar charge vanishes, the geometry reduces to the flat Minkowski spacetime. 
	
	Using this seed solution, we then construct its disformed version which provides an exact solution of Petrov type I for a specific shift-symmetric Horndeski theory. Since the scalar field remains unaffected by the field redefinition, this trick enables one to obtain a geometry which describes the GWs generated by the very same scalar pulse within a higher-order scalar-tensor theory. The resulting solution thus offers a simple testbed to analyse the nonlinear phenomenology triggered by the higher-order terms responsible for the scalar-tensor mixing. In the exact solution, the effects of these scalar-tensor mixing are repackaged and controlled by the disformal parameter $B_0$.  Remarkably, while the scalar monopole only generates longitudinal waves in GR, we show that the disformal version contains a nonlinear superposition of both longitudinal and tensorial waves. This shear disappears when i) the scalar charge vanishes and/or ii) when the disformal parameter vanishes, showing that it builds up from the complicated scalar-tensor mixing inherent to the Horndeski theory. We show that the presence of this shear is a highly nonlinear effect, which cannot be accounted for by a perturbative description of the solution. The generation of this shear from a purely scalar monopole provides the main result of this work.  A more formal investigation of the generation of these tensorial waves from a disformal transformation is presented in the companion paper \cite{letter}.
	
	To further confirm this finding, we analyse the polarization of the propagating waves by numerically solving the geodesic equation and constructing the Penrose limit of our radiative solution. This procedure allows one to associate a pp-wave geometry, to a given (lightlike) observer, which encodes the leading local tidal effects of the initial spacetime. The polarization of the GWs can then be read off form the profile of this pp-wave geometry. Using this trick, we show that the pp-wave indeed contains a trace (or breathing) mode and a spin-2 polarizations. We finally use this pp-wave approximation of our radiative solution to compute the velocity memory effects induced by the passage of the pulse on two nearby lightlike test particles, demonstrating the constant shift in the relative velocity inherited by the two particles. In the end, this new exact solution should be considered as a concrete theoretical example enabling one to reveal new surprising effects emerging in the nonlinear radiative regime of a higher-order scalar-tensor theory. We will discuss the possible implications of our results in the conclusion.
	
	This work is organized as follows. In section~\ref{A}, we describe the seed solution and analyze its properties. In subsection~\ref{B}, we review the disformal solution-generating map and present the target theory consisting in a shift-symmetric Horndeski theory. Subsection~\ref{C} presents the new exact radiative solution. We discuss the GWs propagating in this new geometry and present the main result in subsection \ref{sub-D}. Section~\ref{D} is devoted to the analysis of the polarization at the nonlinear level, and to the derivation of the velocity memory effect using the Penrose limit. Finally, we conclude in section~\ref{disc} by a discussion on our results and the perspectives it opens. Some technicalities are presented in the appendices \ref{app:resolution-P}, \ref{App-D}, and \ref{app:trace-polar}.

	\section{Robinson-Trautman solution with a scalar hair} 
	
	\label{A}
	
	In this section, we review the exact solution of the Einstein-Scalar massless system introduced in \cite{Tahamtan:2015sra} which describes a Robinson-Trautman geometry with a scalar hair.  This solution will serve as a seed to construct the non-perturbative exact solution in the Horndeski theory in the next sections. Therefore, we shall first analyze in detail the properties of this geometry and the nature of GWs it contains. Doing so, we shall confirm and improve in several ways the analysis presented initially in \cite{Tahamtan:2015sra, Tahamtan:2016fur}.
	
	%\subsection{Exact solution of the Enstein-Scalar system}
	Consider the Einstein-Scalar system with the action
	\be
	\label{EinS}
	S[g_{\mu\nu}, \phi] = \frac{1}{2} \int \dd^4{x} \sqrt{|g|} \left( \cR - g^{\mu\nu}\partial_{\mu} \phi \partial_{\nu} \phi \right) \,,
	\ee
	where we work in the unit $M_{\rm Pl}=(8\pi{G})^{-1/2}=1$ and metric signature $(-,+,+,+)$. Taking variation of the action w.r.t. the metric and the scalar field, the field equations take the standard form
	\begin{align}
		\cR_{\mu\nu} - \frac{1}{2} g_{\mu\nu} \cR & = \cT_{\mu\nu} \;, \\
		\label{EoM-SF}
		\Box \phi & =0 \,,
	\end{align}
	where the energy-momentum tensor is given by
	\be\label{EMT-SF}
	\cT_{\mu\nu} = \phi_{\mu} \phi_{\nu} - \frac{1}{2} g_{\mu\nu} \left(g^{\alpha\beta} \phi_{\alpha} \phi_{\beta}\right)  \,,
	\ee
	with $\phi_\mu \equiv \nabla_\mu\phi$.
	As shown in \cite{Tahamtan:2015sra}, the above field equations admit the following radiative exact solution given by
	\begin{align}\label{BG-RT}
		\rd s^2 & =  - \frac{ r \partial_u U + K(x,y) }{U(u)}  \rd u^2 - 2 \rd u \rd r + \frac{r^2 U^2(u)- C^2_0}{U(u)P^2(x,y)} (\rd x^2 + \rd y^2)  \,,\\
		\label{BG-SF}
		\phi (u, r) & = \frac{1}{\sqrt{2}} \log{\left[ \frac{r U(u) - C_0}{r U(u)+ C_0}\right]} \,;
		\hspace{1.5cm}
		U(u) = \gamma e^{\ic^2 u^2 + \eta u} \,,
	\end{align}
	where $(\ic, \eta, C_0, \gamma)$ are constants of integrations and $P(x,y)$ and $K(x,y)$ should satisfy
	\begin{align}
	\label{eq:ang-K}
	K(x,y) = \Delta \log{P}(x,y) \,, \qq{and} \Delta K(x,y)  = 4 C^2_0 \ic^2 \equiv \alpha^2 \,,
	\end{align}
	with
	\be
	\Delta \equiv P^2 (\partial^2_{xx} + \partial^2_{yy}) \,.
	\ee
	At this level, the solution is parametrized by the set of four independent constants $(\ic, \eta, C_0, \gamma)$. The scalar charge is encodes in $C_0$, i.e. when $C_0=0$, the scalar field vanishes.\footnote{More precisely, the equation of motion for the scalar field \eqref{EoM-SF} can be written as $\nabla_\mu J^\mu = 0$, where $J^\mu\equiv\phi^\mu$ is the Noether current associated with the shift-symmetry. It can be easily seen that $J^\mu=0$ for $C_0=0$.} The parameter $\eta$ only affects the position of the maximum of the function $U(u)$ and it can be reabsorb by a suitable rescaling of the $u$-coordinate. Therefore, without loss of generality, we set $\eta=0$. For now on, we shall consider the case where $\gamma>0$, $\ic >0$ and $C_0 > 0$.\footnote{Note that we have excluded the case $C_0=0$. In this case, the energy-momentum tensor \eqref{EMT-SF} vanishes for the background configurations \eqref{BG-RT} and \eqref{BG-SF} and one finds the stealth Robinson-Trautman geometry: the GR Robinson-Trautman solution with a non-propagating scalar field. Stealth solutions are usually strongly coupled and we exclude this case from our consideration. It is worth mentioning that the higher-order operators may naturally resolve this issue in the context of the so-called Scordatura mechanism \cite{Motohashi:2019ymr,		Gorji:2020bfl,Gorji:2021isn,DeFelice:2022xvq}.} The range of the coordinates will be discussed in the next subsection when investigating the singularities.

\begin{figure}[!htb]
		\centering
		\includegraphics{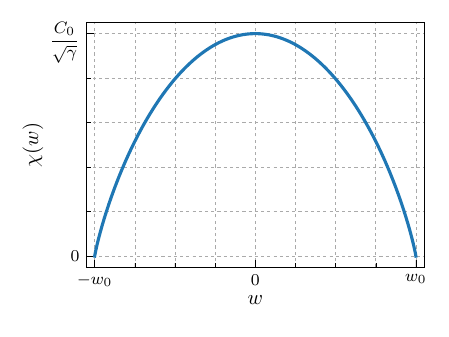}
		\caption{The function $\chi(w)$ is defined in the interval $[-w_0, w_0]$ and takes values in the interval $[0, C_0/\sqrt{\gamma}]$.}
		\label{fig:chi}
	\end{figure}

	It is convenient to switch from the coordinates $(u,r, x,y)$ to a new set of coordinates $(w, \rho, x,y)$ given by
	\begin{align}
		\dd{w} = \frac{\rd u}{\sqrt{U}}\,, \qquad \text{and} \qquad \rho = r \sqrt{U}  \,.
	\end{align}
	The first relation cannot be integrated into a simple analytic form. However, if one uses the error function $\text{erf}$ defined by $\text{erf}(x) = (2/\sqrt{\pi}) \int_0^{x} \exp[-t^2] \dd{t}$ and its inverse $\text{erf}^{-1}$, then one has
	\begin{equation}
		w = w_0 \;\text{erf}\left(\frac{u \ic}{\sqrt{2}}\right) \,,
	\end{equation}
	which implies
	\be
	u = \frac{\sqrt{2}}{\ic} \text{erf}^{-1}\left(\frac{w}{w_0}\right) \,, \qq{and} w_0 = \frac{\sqrt{\pi}}{\ic\sqrt{2\gamma}} \,.
	\ee
	In these coordinates, null past infinity is cast to $w = -w_0$ while null future infinity is cast to $w = +w_0$. 
	Furthermore, the line element and the scalar profile take the simpler form
	\begin{align}
		\label{met}
		\rd s^2 & = - K(x,y) \rd w^2 - 2 \rd w \rd \rho + \frac{\rho^2 - \chi^2(w)}{P^2(x,y)}(\rd x^2 + \rd y^2) \,, \\
		\label{scalprof}
		\phi(w,\rho) & = \frac{1}{\sqrt{2}} \log{\left[ \frac{\rho - \chi(w)}{\rho+ \chi(w)}\right]} \,;
		\hspace{1cm} \chi (w)= \frac{C_0}{\sqrt{U(w)}} \,.
	\end{align}
	The function $\chi(w)$ is depicted in Fig.~\ref{fig:chi}. Explicitly, one has $\chi(w\rightarrow \pm w_0) \rightarrow0$ and $\chi(0) =C_0/\sqrt{\gamma}$ such that $\chi$ remains a bounded function taking values in the range $[0, C_0/\sqrt{\gamma}]$. Finally, the kinetic energy of the scalar field reads
	\begin{align}
		\label{X}
		X \equiv g^{\alpha\beta} \phi_{\alpha} \phi_{\beta}   =  \frac{2\chi}{(\rho^2-\chi^2)^2} \left( \chi K(x,y) +2 \rho \chi'\right) \,,
	\end{align}
	where a prime denotes a derivative w.r.t. the $w$-coordinate.
	We stress that a second branch of solution can be obtained where the function $\chi(w)$ diverges in the regimes $w\rightarrow \pm w_0$ and vanishes at $w=0$. However, we discard this branch because one can show that the kinetic energy of the scalar field diverges in this case, leading to a pathological behavior. 
\subsection{Behavior of the solution}
	
	The effect of the scalar field can be understood by looking at the equation (\ref{eq:ang-K}) satisfied by $K(x,y)$. For $\alpha=0$, equation (\ref{eq:ang-K}) has the solution 
	\be
	\label{P0}
	P_0 = \frac{1 + x^2 + y^2}{2} \,, \qquad K_0=1 \,.
	\ee
	which is nothing but the geometry of a two-sphere in Cartesian coordinates. Thus, the deviation from the spherical symmetry is controlled by the parameter $\alpha$. In our setup, the spherical symmetric solution $\alpha = 2\ic C_0 = 0$ can be achieved in two different ways. The first possibility is when $\ic =0$ but $C_0 \neq 0$ such that the geometry reduces to a spherically symmetric time-dependent geometry with a scalar hair. The second possibility is $C_0=0$ when the geometry reduces to Minkowski spacetime $\lim_{C_0 \to 0} \rd s^2 = - \rd w^2 + 2 \rd w \rd u + \rho^2 \rd \Omega^2$ in agreement with the Birkoff's theorem. In particular, this underlies the fact that there is no mass contribution in this solution, and that the only source of curvature is the scalar monopole. However, as we have already mentioned, the case $C_0=0$ provides a stealth solution which can be problematic. We thus focus on the case $\alpha\neq0$ with $\ic >0$ and $C_0 >0$. In this case, the two-dimensional space $\cB$ spanned by $(\partial_x,\partial_y)$ has a non-homogeneous time-dependent two-dimensional curvature. As we shall see, it corresponds to a GW pulse induced by the scalar field which we shall analyze in the following. 
	
\begin{figure}[!htb]
		\centering
		\includegraphics{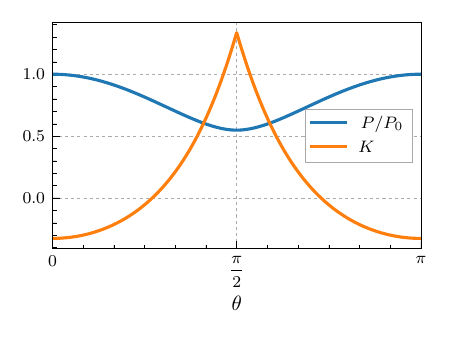}
		\caption{Plots of $\flatfrac{P}{P_0}$ and $K$ obtained numerically as functions of the azimuthal angle $\theta$.}
		\label{fig:K}
	\end{figure}

	A first analysis of the singularities and the apparent horizons with $\alpha\neq 0$ has been presented in \cite{Tahamtan:2015sra}. However, most of the expressions provided by the authors remained implicit as they did not provide an explicit solution for the profile of the function $K(x,y)$. Here, we provide an explicit solution for 
	$K(x,y)$ by solving numerically the equation (\ref{eq:ang-K}). The details of the numerical solution is summarized in appendix~\ref{app:resolution-P} and the profile of both $K(x,y)$ and $P(x,y)$ are depicted\footnote{For clarity, we use in the plots the spherical coordinates $(\theta, \varphi)$ on the sphere, defined in Eq.~\eqref{cood}.} in Fig.~\ref{fig:K}. Notice that while $P(x,y)$ is ${\cal C}^2$, the function $K(x,y)$ is only ${\cal C}^0$ but not ${\cal C}^1$. It follows that the two-dimensional space $\cB$ spanned by $\partial_x$ and $\partial_y$ has a positive curvature peaked at the equator. As we shall see, the present solution can also be interpreted as an impulsive scalar wave.  In the following, we use this numerical solution for the function $K(x,y)$ to discuss the behavior of Tahamtan and Svitec's solution, thus improving on the analysis reported in \cite{Tahamtan:2015sra, Tahamtan:2016fur}.
	
	Now, we can analyze the singularities and the apparent horizons in this geometry. The curvature invariants are given by  
	\be
	{\cal R}  = X \,, \qq{and} \cK = {\cal R}_{\alpha\beta\mu\nu} {\cal R}^{\alpha\beta\mu\nu}= 3 {\cal R}_{\mu\nu} {\cal R}^{\mu\nu} = 3 {\cal R}^2 \,.
	\ee
	Thus the expression (\ref{X}) for $X$ reveals a singular hypersurface located at $
	\rho_s = \chi(w)$. It follows that the singularity is located on the hypersurface $\rho_s=0$ at $w= -w_0$. This hypersurface expands up to $\rho_s=C_0/\sqrt{\gamma}$ before contracting again to reach its initial position $\rho_s=0$ at $w= w_0$.
	Notice that the presence of this singularity restricts the range of the $\rho$-coordinate to $\rho > \chi(w)$ while $w\in [- w_0, + w_0]$ .
	
	To locate the possible apparent horizons, we introduce the null tetrad
	\begin{align}
		\ell^{\mu}\partial_{\mu} & = \partial_\rho \,, \\
		n^{\mu} \partial_{\mu} & = \partial_w + \frac{P(x,y)^2}{\rho^2 - \chi^2} (M_x \partial_x + M_y \partial_y)  
		- \frac{1}{2} \left[ K(x,y) - \frac{P(x,y)^2}{\rho^2 - \chi^2} \left(M_x^2 + M_y^2\right)\right] \partial_\rho \,, \\
		m^{\mu} \partial_{\mu} & = \frac{P(x,y)}{\sqrt{2(\rho^2 - \chi^2)}} \left[ (M_x + i M_y)\partial_\rho + \partial_x + i \partial_y \right] \,.
	\end{align}
	which is orthogonal to the surface $\rho = M(x, y)$, i.e. such that $g^{\mu\nu} \ell_\mu \partial_\nu M = 0$ and which satisfies the standard orthogonality relations
	\be
	\ell^{\mu} n_{\mu} = -1 \,, \qquad m^{\mu} \bar{m}_{\mu} = 1 \,, 
	\ee
	while $\ell^\mu m_\mu = n^\mu m_\mu = 0$.
	The null vectors $\ell$ and $n$ correspond respectively to out-going and in-going null rays.
	The expansions of these vectors are given by
	\begin{align}
		\Theta_{\ell} &%= -\Re(\rho_\mathrm{NP}) 
		= \frac{\rho}{\rho^2 - \chi^2}\,,\\
		\Theta_n &%= \Re(\mu_\mathrm{NP}) 
		= \frac{ \Delta_S M-\rho K(x,y) - 2\chi(w)\chi'(w)}{2(\rho^2 - \chi^2)}  - \frac{\rho \|\nabla_S M \|^2}{2(\rho^2 - \chi^2)^2} \,.
	\end{align}
	Thus $\Theta_{\ell}$ cannot vanish. However, the expansion for the ingoing vector $n^{\mu}\partial_{\mu}$ vanishes, i.e. $\Theta_n = 0$, on the null hypersurface $\rho=M(x, y)$ when the function $M(x,y)$ satisfies the following equation
	\begin{align}
		\label{hor}
		\Delta M -M(x, y) K(x,y)  - 2\chi(w)\chi'(w)  -  \frac{\rho \|\nabla_S M\|^2}{M(x, y)^2 - \chi(w)^2}  =0 \,.
	\end{align}
	Provided this equation is satisfied, the hypersurface $\rho =M(x,y)$ corresponds to a null apparent trapping horizon. Thus, this exact solution corresponds to a singular radiative geometry with a propagating apparent horizon. The exact motion of this null hypersurface is in general too complicated to be integrated, even numerically. We stress that this challenge in investigating the motion of the apparent horizon is a general difficulty one encounters in this type of radiative solutions. See \cite{Tod, Chow:1995va, deOliveira:2009mc} for related investigations for Robinson-Trautman geometries. We shall also recover this limitation when investigating the disformed solution. Now, we would like to gain intuition on the GWs propagating in this geometry. 
	
	\subsection{Longitudinal wave pulse}
	
	We now show that this solution can be understood as a scalar pulse carrying finite energy.   %at the form of the function $\chi(w)$ depicted in Figure~\ref{fig:chi}.
	
	\subsubsection{Asymptotic regimes}
	To understand the waveform propagating to $\cI^{+}$, we shall describe the different regimes of interest.
	\begin{itemize}
		\item In the remote past and far future ($w \rightarrow \pm w_0$), one has $\chi(w) \rightarrow 0$, such that the kinetic term \eqref{X} vanishes:  $X=0$. Moreover, the metric and the scalar field behave as follow:
		\begin{align}
			\label{asymmet}
			&\begin{multlined}
				\lim_{w \to \pm w_0}  \rd s^2 = - K(x,y) \rd w^2 - 2 \rd w \rd \rho + \frac{\rho^2 }{P^2(x,y)}(\rd x^2 + \rd y^2) \,,\end{multlined}\\
			&\lim_{w \to \pm w_0}  \phi =  \frac{\sqrt{2} \chi(w)}{\rho}  \rightarrow 0 \,.
		\end{align}
		Thus, except at $\rho=0$ where the singularity lies, the scalar field and its kinetic energy vanish everywhere in these two asymptotic regimes. The metric is static and non-spherically symmetric since the term $P(x,y)$ encodes the deformation w.r.t. the unit two-sphere (corresponding to $P = P_0$). \\
		\item At the maximum of the pulse, $w=0$, the kinetic term is given by
		\be
		X = \frac{2 \gamma C^2_0 }{(\gamma \rho^2 - C^2_0)^2} K(x,y) \,,
		\ee
		which is positive, while the metric and the scalar field become
		\begin{align}
			\label{asymscal}
			&\begin{multlined}\lim_{w \to 0}  \rd s^2 = - K(x,y) \rd w^2 - 2 \rd w \rd \rho + \frac{\gamma \rho^2- C^2_0 }{\gamma P^2(x,y)}(\rd x^2 + \rd y^2) \,, \end{multlined}\nn \\
			&\lim_{w \to 0}  \phi = \frac{1}{\sqrt{2}} \log{\left[ \frac{\sqrt{\gamma}\rho - C_0}{\sqrt{\gamma}\rho+ C_0}\right]} \,.
		\end{align}
		Thus, the scalar pulse which is built along time acts as a Gaussian scaling of the two-dimensional space $\cB$ spanned by $(\partial_x, \partial_y)$. After increasing, the rescaling tends to zero, giving back the initial form of the metric. 
	\end{itemize}
	To see this, it is useful to compute the curvature $\cK$ of $\cB$ whose induced metric is
	\be
	\rd s^2_{\cB} =  \frac{\rho^2 - \chi^2}{P^2(x,y)}(\rd x^2 + \rd y^2) \,.
	\ee
	Its two-dimensional curvature takes the form
	\be
	\cK(w,x,y) = \frac{1}{C^2_0} \chi^2(w) K(x,y) \,,
	\ee
	which shows the rescaling of $K(x,y)$ by the function $\chi(w)$. Indeed, at $w\rightarrow \pm w_0$, one has $\cK =0$ while at $w=0$, the curvature is $\cK(x,y) = K(x,y)/\gamma$.
	Notice that such a GW does not provide any deformation in the direction orthogonal to the direction of propagation. The only deformation takes place along the direction of propagation, providing an explicit example of a longitudinal GW induced by the scalar monopole (\ref{scalprof}). 
	
	\subsubsection{Optical scalars}
	
	To further understand the nature of the nonlinear GW, it is useful to consider its effect on a congruence of null geodesics. To that end, let us introduce the simplified tetrad:
	\begin{align}
		\label{seednullvec1}
		\ell^{\mu}\partial_{\mu} & = \partial_\rho \,, \\
		n^{\mu} \partial_{\mu} & = \partial_w - \frac{1}{2} K(x,y) \partial_\rho \,, \\
		\label{seednullvec3}
		m^{\mu} \partial_{\mu} & = \frac{P(x,y)}{\sqrt{2(\rho^2 - \chi^2)}} \left( \partial_x + i \partial_y \right) \,.
	\end{align}
	The null vectors $\ell$ and $n$ correspond respectively to out-going and in-going null rays. The vector $\ell$ is geodesic, i.e. $\ell^{\alpha}\nabla_{\alpha} \ell^{\mu} =0$. The associated geodesic motion can be easily integrated for the associated lightlike observer. Introducing an affine parameter $\lambda$, the solution to the geodesic equation with $\ell^{\mu} \partial_{\mu}$ being the tangent vector is given by
	\begin{align}
		\label{eq:geodesic-nodisf}
		\rho(\lambda) = \lambda + \rho_0 \,, \qquad  w(\lambda) = w_0 \,, \qquad  x(\lambda) = x_0 \,, \qquad   y(\lambda) =y_0 \,,
	\end{align}
	where $(w_0, \rho_0, x_0, y_0)$ are the initial conditions. Thus the geodesic describes a photon traveling in the radial direction without any motion on the two-sphere.
	As expected, the Sachs optical parameters $( \Theta, \omega, \sigma)$ describing the expansion $\Theta$, twist $\omega$ and shear $\sigma$ of this null congruence, which are defined in \eqref{sigma-omega-theta}, are given by
	\begin{align}
		\Theta (\rho, w) = - \frac{\rho}{\rho^2 - \chi^2} \,, \quad \omega = 0 \,, \qq{and}  \sigma  = 0 \,.
	\end{align}
	This shows that the only effect induced by the GW consists in an expansion of the congruence, i.e. confirming the purely longitudinal nature of the wave. This exact solution thus belongs to the Robinson-Trautman family of geometries admitting a twist-free, shear-free but expanding null congruence. See \cite{Podolsky:2016sff} for details on the properties of Robinson-Trautman geometries. 
	
	\subsubsection{Petrov type}
	
	We can compute the Weyl scalars to identify the Petrov type of this geometry. One can show that the speciality index (defined in appendix~{\ref{App-D}}) is given by
	\be
	S =1 \,,
	\ee 
	demonstrating that the geometry is algebraically special. Furthermore, one can consider the quantities $(L,M,  I)$ also defined in appendix~\ref{App-D} and show that $N \neq 0$ and $M\neq 0$ when $K(x,y)$ is not constant, while $N=M=0$ when $K(x,y) =1$. This demonstrates that this spacetime is of Petrov type II when $K(x,y)$ is not a constant, while it is of type D when $K(x,y) =1$, i.e. in the spherically symmetric case. See \cite{Bini:2021aze} for details on the Petrov classification. This is consistent with previous analysis reported in \cite{Podolsky:2016sff}. It is instructive to provide the expressions of the non-vanishing Weyl scalars for our choice of null directions. They read
\begin{align}\label{Psis-seed}
\Psi_2 = \frac{1}{6} X \,, 
\qquad
\Psi_3  = -\frac{i \rho P \partial_z K}{2
	\left(\rho^2-\chi^2\right)^{3/2}} \,,
\qquad
\Psi_4 =-\frac{P^2}{2(\rho^2-\chi^2)}
\left(  \partial_z^2 K + 2 \partial_z K \partial_z \ln{P}
\right) \,,
\end{align}
where we have presented the results in terms of the complex coordinate $z=(x+iy)/\sqrt{2}$. 

\begin{figure}[!htb]
		\centering
		\includegraphics{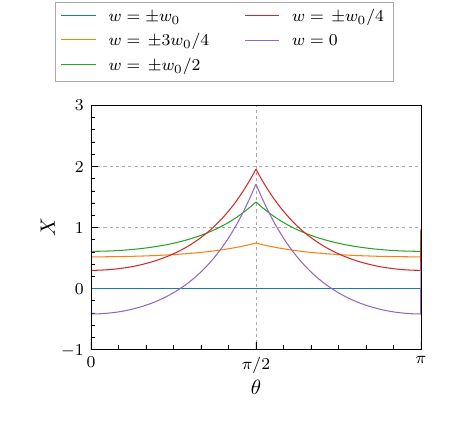}
		\caption{The kinetic energy $X$ of the scalar field is plotted for $\alpha = \alpha_0$, $C_0 = 1$, $\gamma = 1$ and $\rho = 1.5$.}
		\label{pt_Xorig}
	\end{figure}

Notice that $\Psi_2$ is proportional to the kinetic energy of the scalar field $X$. Since it provides the key quantity for the matter sector, we plot its behavior in Fig.~\ref{pt_Xorig}. Its behavior confirms the interpretation of the solution as a scalar energy pulse. Notice also that while $X$ is positive almost everywhere, it admits a negative value near the two poles at $w=0$. This implies that while the gradient of the scalar field is spacelike almost everywhere (for our choice of sign in the definition of $X$), it becomes time-like at the maximum of the pulse in a small region near the two poles. Whether the motion of apparent horizon located at $\Theta_n =0$ is responsible for this behavior is an open question as we were not able to integrate its time-development encoded in \eqref{hor}. Answering this question would require a complete investigation of the motion of the horizon which goes beyond the scope of this work.
	
	%We first notice that $\Psi_2$ and $\Psi_4$ correspond to a pulse picked on and symmetric w.r.t.  the equatorial line $\theta= \pi/2$, while $\Psi_3$ has growing value at both the north and south poles. 
	%In the asymptotic regime, when $\omega=\pm w_0$, we observe that $\Psi_2$ vanishes, in agreement with the asymptotic form of the metric (\ref{asymscal}). Moreover, it reflects the fact that in this geometry, there is no mass term sourcing the curvature but only a scalar monopole. The pulse centered at the equator grows in time up to some $\omega_{\ast}$ (different for each Weyl scalars) before decreasing back to its original value. 
	%It provides a concrete representation of the time-dependent curvature of the topological two sphere in this radiative geometry, confirming the picture of a longitudinal scalar wave rescaling the induced geometry of the 2d space $\cB$. 
	Let us emphasize that the representation of the waveform is made possible here by having first solved numerically the profile for the function $K(x,y)$, a step which was missing in \cite{Tahamtan:2015sra, Tahamtan:2016fur}. In that sense, the representation of the scalar pulse described above provides a first result of this work. Equipped with this exact solution, we can now present the new exact radiative solution in Horndeski gravity and compare its nonlinear phenomenology with its GR counterpart.
	
	%{\Ji Remark: the e.o.m are second order and the solution is thus symmetric under the change $w \rightarrow - w$. Therefore, should we not just consider half of the time evolution, corresponding to the range $[-w_0, 0]$ and $[0, w_0]$ ? Then, the next question is what is the scalar profile at $w = - w_0$ and at $\rho = \chi(-w_0) =0$, i.e. at the singularity ? It seems that it tends to $0$ at fixed $\rho$ but it diverges at the singularity.}
	
	\section{Non-perturbative GW solution in Horndeski gravity}

	\subsection{Disformal mapping}\label{B}
	
	In order to construct this exact solution in Horndeski gravity, we shall use the disformal solution-generating method \cite{BenAchour:2020wiw}. While the method applies to any scalar-tensor theory, it will be enough to consider the Einstein-Scalar action $S[g_{\mu\nu}, \phi]$ of \eqref{EinS} as a seed theory for our purposes. Performing a disformal mapping on the metric $g_{\mu\nu}$ given by
	\be
	\label{dis}
	\left( g_{\mu\nu}, \phi \right) \rightarrow \left( \tilde{g}_{\mu\nu} = g_{\mu\nu} + B_0 \phi_{\mu} \phi_{\nu}, \phi \right) \,,
	\ee
	where $B_0$ is a constant parametrizing the deviation w.r.t. the seed theory, the Einstein-Scalar action $S[g_{\mu\nu}, \phi]$ given by \eqref{EinS} gets corrected by higher-order scalar-tensor contributions giving rise to a new scalar-tensor theory $\tilde{S}[\tilde{g}_{\mu\nu}, \phi]$. This new theory belongs to the shift-symmetric Horndeski family whose action reads
	\begin{align}\label{action-Horndeski}
		\tilde{S}[\tilde{g}_{\mu\nu}, \phi] &  = \int \rd^4 x \sqrt{|g|}\;  \cL_h(\tilde{g}_{\mu\nu}, \phi) \,,%\left[ G_2(X) + G_4(\tilde{X}) \cR  \right. \nn \\
		%&\qquad \qquad   \left. - \; 2G_{4X}(\tilde{X}) \;\left( (\Box \phi)^2 - \phi_{\mu\nu} \phi^{\mu\nu} \right)\right] \;\;\;\;\qquad \;\;
	\end{align}
	with $\cL_h$ the reduced Horndeski Lagrangian given by
	\be
	\label{Horn}
	\cL_h = G_2(\tilde{X}) + G_4(\tilde{X}) \cR - 2G_{4X}(\tilde{X}) \left[ (\Box \phi)^2 - \phi_{\mu\nu} \phi^{\mu\nu} \right] \,,
	\ee
	%where
	%\be
	%\cL_1 = (\Box \phi)^2 - \phi_{\mu\nu} \phi^{\mu\nu}  \qquad \cL_4 =  \phi_{\mu\nu} \phi^{\nu}{}_{\rho} \phi^{\mu} \phi^{\rho}
	%\ee
	and the two functions $(G_2, G_4)$ given by
	\begin{align}
		G_2(\tilde{X})  = - \frac{1}{2} \tilde{X} \,, \qq{and} G_4(\tilde{X})  = \frac{1}{2\sqrt{1-B_0 \tilde{X}}} \,.  
		% \cA(\tilde{X}) & = \frac{B_0}{\sqrt{1-B_0 \tilde{X}}} 
		%  A_4(\tilde{X}) & = - \frac{B^2_0}{\sqrt{1- B_0 \tilde{X}}}
	\end{align}
	See \cite{Kobayashi:2019hrl} for a review on the phenomenology of the Horndeski family of theories. 
	This transformation agrees with the known result that starting from a Horndeski theory (i.e. GR with a minimally coupled massless scalar field in our case), we remain in the Horndeski family after a disformal transformation \cite{BenAchour:2016cay}.
	Under this transformation, the scalar kinetic term $X = g^{\mu\nu} \phi_{\mu} \phi_{\nu}$ is modified. The relations between the new kinetic term $\tilde{X}= \tilde{g}^{\mu\nu} \phi_{\mu} \phi_{\nu}$ and the seed one $X$ are given by
	\be
	\tilde{X} = \frac{X}{1+B_0 X}\,, \qquad X = \frac{\tilde{X}}{1-B_0 \tilde{X}} \,.
	\ee
	At the level of the solution space, a given exact solution to the Einstein-Scalar system $(g_{\mu\nu}, \phi)$ is mapped to another exact solution $(\tilde{g}_{\mu\nu}, \phi)$ of the Horndeski theory. Let us emphasize that at this level, the disformal transformation is a pure field redefinition and does not contain any new physics. \textit{The key point enters when choosing to which metric the matters gets minimally coupled.} Indeed, for test fields minimally coupled to the disformal metric $\tilde{g}_{\mu\nu}$, the behavior will be quite different from test fields minimally coupled to the seed metric $g_{\mu\nu}$. Thus, this simple method allows one to construct exact solutions of degenerate higher-order scalar tensor theories without having to solve for their complicated field equations. When combined to the Petrov classification, it provides a powerful tool to organize the construction of new solutions \cite{Achour:2021pla}. See \cite{Faraoni:2021gdl, Bakopoulos:2022csr, BenAchour:2020fgy, Anson:2020trg} for the construction of new black hole solutions beyond the stealth sector in DHOST gravity, among which exact rotating black holes solutions \cite{BenAchour:2020fgy, Anson:2020trg,Baake:2021kyg}. 
	
	Another interesting feature of the disformal transformation (\ref{dis}) is that the scalar profile $\phi$ remains unchanged. Thus, it allows to contemplate, for a given scalar profile, the modification on the metric sector induced by the higher-order terms in the action $\tilde{S}[\tilde{g}_{\mu\nu}, \phi]$. The parameter $B_0$ can be interpreted as a dimensionless coupling constant encoding the strength of the higher-order modifications. Having clarified the main properties of the method, we are now ready to construct the new radiative solution of the specific Horndeski theory.
	
	\subsection{Exact radiative solution}\label{C}
	
\begin{figure}[!htb]
		\centering
		\includegraphics{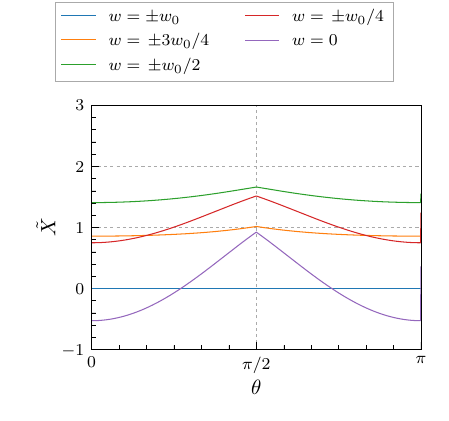}
		\caption{The kinetic energy $\tilde{X}$ of the scalar field after disformal transformation, for $\alpha = \alpha_0$, $C_0 = 1$, $\gamma = 1$ and $\rho = 1.5$.}
		\label{plot_X}
	\end{figure}

	As emphasized previously, the scalar profile remains the same such that
	\begin{align}
		\label{scalproff}
		\phi(w,\rho) = \frac{1}{\sqrt{2}} \log{\left[ \frac{\rho - \chi(w)}{\rho+ \chi(w)}\right]} \,;
		\qquad
			\chi(w) =  \frac{C_0}{\sqrt{\gamma}} \exp\left[\left(\text{erf}^{-1}\left[w/w_0\right]\right)^2\right]  \,.
	\end{align}
	Notice that since the scalar field profile does not depend on the angular coordinates $(x,y)$, it can be again considered as a pure scalar monopole.
	The new exact solution is given by
	\begin{align}
		\label{metdis}
		\rd s^2 & = - K(x,y)  \rd w^2 - 2 \rd w \rd \rho + \frac{\rho^2 - \chi^2(w)}{P^2(x,y)}(\rd x^2 + \rd y^2)  \nn \\
		& \;\;\; + B_0 \left[ \phi^2_w  \;  \rd w^2 + 2 \phi_w \; \phi_{\rho} \;\rd w \rd \rho + \phi^2_{\rho} \; \rd \rho^2 \right] \,.
	\end{align}
	The first line corresponds to the original metric and the functions $K(x,y)$ and $P(x,y)$ remain unchanged, i.e. they are given by
	\begin{align}
		K(x,y) = \Delta \log{P}(x,y) \,, 
		\qquad
		\Delta K(x,y) = 4 C^2_0 \ic^2 = \alpha^2 \,,
	\end{align}
	where, similar to the seed metric, $C_0$ (or $\alpha$ for $\ic\neq0$) characterizes the scalar charge for the disformed metric. We recall that the functions $\chi(w)$ and $K(x,y)$ are depicted in Fig.~\ref{fig:chi}  and Fig.~\ref{fig:K} respectively. The second line in \eqref{metdis}, that is proportional to the parameter $B_0$, encodes the new effects induced by the higher-order terms.
	These new terms are constructed from the components of the scalar gradient which read explicitly
	\be
	\label{phiw}
	\phi_{\rho} = \frac{2\chi(w)}{\rho^2- \chi^2} \,, \qq{and} \phi_w = \frac{- 2 \rho \chi'(w)}{\rho^2- \chi^2} \,.
	\ee
	It follows that the kinetic energy of the scalar field in this new geometry can be written as
	\be
	\tilde{X}= \frac{\phi_{\rho}(K \phi_{\rho} - 2 \phi_{w} + 2 B_0 \phi_{\rho} \phi^2_{w})}{1 + B_0 \phi_{\rho}(K \phi_{\rho} + 2 \phi_{w})} \,,
	\ee
	which is depicted in Fig.~\ref{plot_X}. One can notice that it has the same qualitative behavior as for the seed solution.
	We emphasize that this geometry is by construction a non-perturbative and exact solution of the Horndeski model given by the action \eqref{action-Horndeski}. Now, our goal is to analyze the phenomenology of this exact solution. 
	
	Computing the Kretschmann invariant, one finds that the GR singularity at $\rho = \chi(w)$ is preserved  while additional singularities show up on the hypersurface defined by
	\be\label{singularity}
	\rho^4 - 2 \left( \rho^2 - B_0 K(x,y)\right) \chi^2 + \chi^4 + 4 B_0 \rho \chi \chi' =0 \,.
	\ee
	These new singularities can be analyzed as follows. When $w\rightarrow \pm w_0$, we have $\chi=0$ and $\chi'$ constant. Thus, in these regimes, the only solution is $\rho_s = \chi(\pm w_0)=0$ and there are no new singularities.  When the pulse reaches its maximum, at $w=0$, one has $\chi(0) = C_0/\sqrt{\gamma}$ and $\chi'(0) =0$ and the new singularity, if it exists, is located at
	\begin{align}
		\rho^2_{\ast} = \frac{C_0^2}{\gamma} \qty(1 \pm \sqrt{\frac{-2B_0 K(x,y) \gamma}{C_0^2}}) \,.
	\end{align}
	Therefore, the disformal transformation generates a new singularity but this additional singular hypersurface remains bounded to $\rho_{\ast}$ when the pulse reaches its maximum. One can also identify the equation for the presence of an apparent trapping horizon but the explicit expression is not particularly illuminating and we shall not reproduce it here. In the end, regarding the locus of the singularities and the possible apparent horizons, the new solution behaves qualitatively the same as the seed one, demonstrating that the disformal transformation has only minimal impact on these properties. We now turn to the main focus, namely the analysis of the waveform in this new geometry.
	
	\subsection{Interpreting the GW}\label{sub-D}
	
	\subsubsection{Asymptotic regimes}
	
	To understand the time-development of the solution, we first investigate the different regimes.
	\begin{itemize}
		\item In the remote past and far future ($w\rightarrow \pm w_0$), the kinetic energy is again vanishing, i.e.  $\tilde{X}=0$. The only component of the scalar gradient which survives is $\phi_{w}$ (given by (\ref{phiw})) such that the metric and the scalar field reduce to
		\begin{align}
			& \lim_{w \to \pm w_0}  \rd s^2  = - \left[K(x,y) - \frac{Q^2}{\rho^2}\right] \rd w^2 - 2 \rd w \rd \rho  + \frac{\rho^2 }{P^2(x,y)}(\rd x^2 + \rd y^2) \,, \\
			& \lim_{w \to \pm w_0}  \phi =  \frac{\sqrt{2} \chi(w)}{\rho}  \rightarrow 0 \,.
		\end{align}
		In both asymptotic regimes, the metric is again static and non-spherically symmetric. However, there is a qualitative difference with the GR solution which shows up in the component $g_{ww}$. This term inherits a new contribution where
		\be
		\label{Q}
		Q \equiv 2 \sqrt{B_0} \lim_{w \to \pm w_0} \chi'(w) \,.
		\ee
		Thus, while the scalar field vanishes everywhere except at the singularity $\rho=0$, its $w$-derivative does not and contributes to the metric. The specific form of this new contribution suggests that it might be considered as an effective electric charge in this regime.
		
		\item At the maximum of the pulse ($w = 0$), the kinetic energy reaches a non-vanishing positive value given by
		\be
		{\tilde X} = \frac{2 \gamma  C^2_0 K(x,y)}{(\gamma\rho^2 - C_0^2)^2 + 2 B_0 \gamma C_0^2 K(x,y) } \,.
		\ee
		The metric and the scalar profile become
		\begin{align}
			& \lim_{w \to 0}  \rd s^2  = - K(x,y) \rd w^2 - 2 \rd w \rd \rho + \frac{4B_0 \gamma C^2_0}{ (\gamma \rho^2-C_0^2)^2} \rd \rho^2  + \frac{\rho^2 }{P^2(x,y)}(\rd x^2 + \rd y^2) \,,\\
			& \lim_{w \to 0}  \phi =  \frac{1}{\sqrt{2}} \log{\left[ \frac{\sqrt{\gamma}\rho - C_0}{\sqrt{\gamma}\rho+ C_0}\right]} \,.
		\end{align}
		Thus when the pulse reaches its maximum, the contribution in $g_{ww}$ present in the asymptotic regimes vanishes since $Q$ is proportional to $\chi'$. On the other hand, the metric inherits another contribution in $g_{\rho\rho}$.
	\end{itemize}
	Now, we can investigate the GWs propagating in this geometry.

	\subsubsection{Null tetrad and Weyl scalars}
	
	To do so, we proceed as follows. We first introduce four null vectors
	\begin{align}
	E^{\mu}_A = (\tilde{\ell}^\mu, \tilde{n}^\mu, \tilde{m}^\mu, \tilde{\bar{m}}^\mu) \,,
	\end{align} 
	with the standard orthogonality relations
	\begin{align}
		\label{nullvec}
		& \tilde{\ell}^{\mu} \tilde{n}_{\mu} = -1 \,, \quad \tilde{m}^{\mu} \tilde{\bar{m}}_{\mu} = 1 \,, \qquad  \tilde{\ell}^\mu \tilde{m}_\mu = \tilde{n}^\mu \tilde{m}_\mu = 0 \,.
	\end{align}
	The choice of null vector basis is not unique since it is defined up to Lorentz transformations. We choose $E^{\mu}_A$ to be the associated null tetrad
	%\be
	%\ell^{\mu} = E^{\mu}_{\lambda} \rd\lambda \;, \qquad n^{\mu} = E^{\mu}_v \rd v \qquad m^{\mu} = E^{\mu}_a \rd X^a 
	%\ee
	%satisfying $g^{\mu\nu} =  E^{\mu}_{\mu} E^{\nu}_B \eta^{AB}$
	constructed such that $\tilde{\ell}^{\mu} = E^{\mu}_{U} \rd U$ corresponds to the tangent vector of a null geodesic $\gamma$ with affine parameter $U$. The remaining vectors are selected such that $E^{\mu}_A$ corresponds to a parallel transported frame (PTF), i.e. such that
	\begin{align}
	\tilde{\ell}^{\mu} \tilde{\nabla}_{\mu} E^{\nu}_{A} =0 \,,
	\end{align}
	where the $\tilde{\nabla}_\mu$ is compatible with the disformed metric $\tilde{g}_{\mu\nu}$. In terms of the spin coefficients that are defined in \eqref{spin-coefficients}, it translates into the conditions: $\tilde{\kappa} = \tilde{\epsilon} = \tilde{\pi} =0$. Physically, it corresponds to the projector onto the family of local inertial frames of an observer following the null geodesic $\gamma$. While we could have worked with a general null tetrad in this section, constructing this PTF will reveal useful later on when investigating the Penrose limit and the memory effect. As we are going to see, the main novelty is that, contrary to the GR one, a PTF cannot be shear-less in the disformed geometry. A detailed derivation of the null tetrad (\ref{nullvec}) and its spin coefficients is presented in the companion paper \cite{letter}.
	
	To simplify the different expressions, we shall assume that $B_0 \ll 1$ and expand w.r.t. the disformal parameter\footnote{In this approximation, the functions entering in the Horndeski action \eqref{action-Horndeski} read
		\be
		G_4 = \frac{1}{2} + \frac{1}{4} B_0 \tilde{X} + \frac{3}{16} B^2_0 \tilde{X}^2 + \cO(B^3_0) \,, \qq{and} G_{4X} = \frac{1}{4} B_0 \left( 1 + \frac{3}{2} B_0 \tilde{X} \right)+ \cO(B^3_0) \,.
		\ee}. This expansion enables us to obtain tractable expressions for the main quantities of interest. Yet, the reader should keep in mind that the solution described and analyzed here remains \text{exact} and \text{non-perturbative}.
	Having clarified that point, we now present the analysis up to quadratic order in $B_0$, i.e. neglecting terms of order $\cO(B^3_0)$. Schematically, the components of the PTF can be written as
	%This can be seen from different perspectives:  computing the Petrov type, computing the spin coefficients for a given tetrad, and finally by analyzing the Penrose limit for a given null geodesic. 
	\begin{align}\label{PTF}
		E^{\mu}_A \partial_{\mu} = \left(^{(0)}\!E^{\mu}_A + B_0 \; ^{(1)}\!E^{\mu}_A + B^2_0 \; ^{(2)}\!E^{\mu}_A  \right) \partial_{\mu} \,.
		%\label{tet1}
		%\ell^{\mu}\partial_{\mu} & = \left(  \ell^{\mu}_0 + B_0\ell^{\mu}_1 + B^2_0\ell^{\mu}_2  \right) \partial_{\mu} \\
		%\label{tet2}
		%n^{\mu} \partial_{\mu} & =  \left(  n^{\mu}_0 + B_0 n^{\mu}_1 + B^2_0 n^{\mu}_2  \right) \partial_{\mu}  \\
		%\label{tet3}
		%m^{\mu} \partial_{\mu} & = \left(  m^{\mu}_0 + B_0 m^{\mu}_1 + B^2_0 m^{\mu}_2  \right) \partial_{\mu} 
	\end{align}
	%The null vector $\ell$ represents the tangent vector to a congruence of a null geodesic $\gamma$ while the vectors $(n, m , \bar{m})$ are constructed such that the null tetrad corresponds to a PTF along the null direction $\ell$, i.e. they satisfy
	%\be
	%D n^{\nu} = D m^{\mu} = D \bar{m}^{\mu} =0
	%\ee 
	%where $D = \ell^{\mu} \nabla_{\mu}$.
	%{\Jib We further restrict our PTF tetrad by demanding a vanishing twist, i.e. $\tilde{\omega} =0$ where $\tilde{\omega}$ is the twist for the disformed metric (see \eqref{sigma-omega-theta} for the definition of the twist). }
	The components $^{(i)}\!E^{\mu}_A$ can be solved exactly order by order in $B_0$ but their expressions are in general involved. %We have performed the computation via the XAct mathematica package which we join to the present work for the interested reader. 
	
\begin{figure}[!htb]
		\centering
		\includegraphics{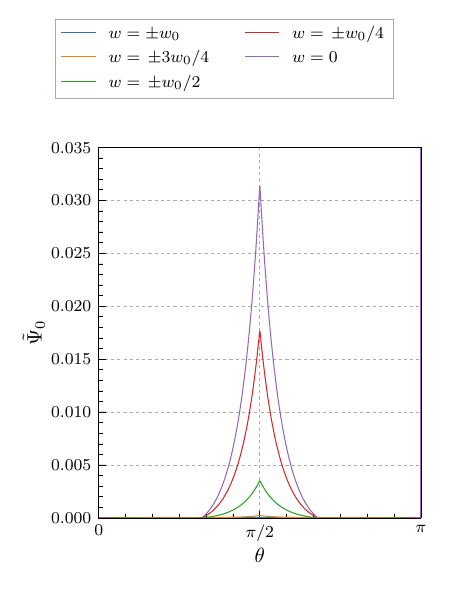}
		\caption{The Weyl scalar $\tilde{\Psi}_0$, for $\alpha = \alpha_0$, $C_0 = 1$, $\gamma = 1$ and $\rho = 1.5$.}
		\label{plot_psi0}
	\end{figure}

	We now can proceed to the nonlinear analysis of the new radiative solution. First, computing the speciality index of this geometry (defined in appendix~\ref{App-D}), one can show that 
	\be
	S = 1 + B^2_0\, G(w,\rho, x,y) + \cO(B^3_0) \,,
	\ee
	where $G$ is a complicated function whose precise form is not relevant. This means that the new solution is not algebraically special, i.e. the geometry is of Petrov type I. Physically, it implies that the Weyl tensor admits four linearly independent principal null directions. While the expressions of the Weyl scalars depend on the choice of frame $E^{\mu}_A$, it will be useful to provide the expression of $\tilde{\Psi}_0$ in the following. For our specific choice of PTF, all the Weyl scalars are non-vanishing and take rather complicated expressions except for the scalar $\tilde{\Psi}_0$ which takes a rather simple form given by
	\begin{align}
		\label{psi0}
		\tilde{\Psi}_0 & = B^2_0 \frac{\chi^4 P \left( P \partial_{\bar{z}\bar{z}} K + 2 \partial_{\bar{z}} K \partial_{\bar{z}} P\right)}{(\rho^2-\chi^2)^5}  + \cO(B^3_0) \,.
	\end{align}
The behavior of $\tilde{\Psi}_0$ at a fixed value of $\rho$ is represented in Fig.~\ref{plot_psi0}. As we shall see, some of the key quantities can be expressed in terms of this scalar $\tilde{\Psi}_0$. 
	
	\subsubsection{Optical scalars}
	
	To further understand qualitatively the nature of the GWs propagating in this geometry, we compute the Sachs optical scalars $(\tilde{\Theta}, \tilde{\omega}, \tilde{\sigma})$ associated to our null tetrad which are defined in \eqref{sigma-omega-theta}. %{\Jib We recall that, by construction, we have chosen a null tetrad such that there is no twist, i.e. $\tilde{\omega} =0$.} 
	Neglecting terms of order $\cO(B^3_0)$, one can write the expansion scalar $\tilde{\Theta}$ as
	\be
	\tilde{\Theta} = \Theta_0 + B_0 \Theta_1 + B^2_0 \Theta_2 + \cO(B^3_0) \,,
	\ee
	and show that
	\begin{align}
		\Theta_0 & = - \frac{\rho}{\rho^2 - \chi^2} \,,\\
		\Theta_1 & = \frac{1}{4 \chi^2 (\rho^2 - \chi^2)^3} \left[ 4 \rho K \chi^4 + 2\chi\chi' \left( \rho^4  + \rho^2 \chi^2 + 2 \chi^4 \right) + \rho \chi' \log{\left( \frac{\rho - \chi(w)}{\rho+ \chi(w)}\right)} \left( \rho^4 - 2 \rho^2 \chi^2 + \chi^4  \right) \right] \,,
	\end{align}
	while the expression for $\Theta_2$ is too complicated to be reproduced here. 
	Additionally, one finds a non-vanishing shear given by
	\begin{align}
		\label{shear}
		\tilde{\sigma} & = B^2_0 \sigma_2 + \cO(B^3_0) \,,
	\end{align}
	where we have explicitly
	\begin{align}
		\sigma_2 & = - \frac{(\rho^2-\chi^2)}{96\chi^7} \left[ 30 \rho^5 \chi - 80 \rho^3 \chi^3 + 66 \rho \chi^5 + 15 (\rho^2-\chi^2)^3 \log{\left( \frac{\rho - \chi(w)}{\rho+ \chi(w)}\right)} \right]  \Psi_0 \,.
	\end{align}
	This analysis reveals that, additionally to the expansion $\tilde{\Theta}$ induced by the scalar wave, a non-vanishing shear $\tilde{\sigma}$ also shows up at the quadratic order in the disformal parameter $B_0$. 
	
	Since the shear is not Lorentz-invariant, it is natural to wonder whether it might not be a gauge artifact of the chosen frame. To answer to this question, one has to pick up a PT frame, i.e. a frame in which the vector $\tilde{\ell}^{\mu} \partial_{\mu}$ is geodesic and in which the remaining three null vectors are parallel transported along $\tilde{\ell}$. At the level of the spin coefficients, this property translates into $\tilde{\kappa} = \text{Re}(\tilde{\epsilon}) = \tilde{\pi} = 0$. This ensures that one is working in the local inertial frame associated to the null geodesic. Starting from a general frame,  one has to use a suitable sequence of Lorentz transformations to construct this PT frame. Proceeding to this construction, one can show that
	\be
	\tilde{\kappa} = \text{Re}(\tilde{\epsilon}) = \tilde{\pi} =0 \,, \qquad \text{implies that} \qquad \tilde{\sigma} \neq 0 \,.
	\ee
	Therefore, one cannot have a shear-free PT frame in this new exact radiative solution. This conclusion is Lorentz-invariant, in the sense that once the PT frame has been constructed, one can show that the residual freedom to perform Lorentz transformation while remaining PT cannot be used to set the shear $\tilde{\sigma}$ to zero (see the companion paper \cite{letter}). However, this residual gauge freedom can be used to set the twist to zero at second order in $B_0$. It follows that there are no shear-less PT frame in the disformed solution, an effect which descends directly from the higher-order term of the Horndeski theory. %Therefore, once such frame is achieved, there is no gauge freedom left to further modify the non-vanishing shear. 
%This is the key difference with the Horndeski radiative solution, as such a frame does not exist in the disformal solution despite having the very same scalar profile.
	
	Thus, the solution also contains a tensorial mode. While small, this effect is qualitatively completely new w.r.t. to the GR solution and can be interpreted as emerging from  the nonlinear mixing between the scalar and tensor sectors in Horndeski theory. Notice that the expression of the shear (\ref{shear}) in terms of $\tilde{\Psi}_0$ is specific to our choice of frame. While one can always rotate the frame to set $\tilde{\Psi}_0 =0$, the shear will remains non-zero and will be expressed in term of other Weyl scalars in this new frame. In the end, the presence of this non-vanishing shear $\tilde{\sigma}$ shows that the higher-order terms in the Horndeski dynamics lead to a rather surprising new phenomenology. Even in the presence of a pure scalar monopole source, the mixing between scalar and tensors sectors allows for the generation of shear. This is the main result of the present analysis.
	
	While our solution is fully nonlinear and non-perturbative in the scalar field profile, it is worth to look at the perturbative description of our solution. Since the key quantity describing the scalar pulse is $\chi(w)$, it is natural to consider the linear regime where $\phi(\rho, w)=-\chi(w)/\sqrt{2}\rho$ for $\chi(w)\ll{1}$. Then, we see that $\tilde{\Psi}_0 \simeq 0$ at the linear regime and it only acquires non-vanishing value at the order ${\cal O}(\chi^4)$. Therefore, the nonlinear effects that we have found would be lost if one linearizes the system from the beginning. For example, the shear $\tilde{\sigma}$ is proportional to $\tilde{\Psi}_0$ and is absent at the linear level. As we shall see, this can also be seen from the polarizations of the gravitational plane wave derived in the next section. Indeed, the two tensorial polarization $H_{+, \times}$ being also given by the real and imaginary part of $\tilde{\Psi}_0$, the same conclusion applies. Finally, let us point that in this linear regime, the effective electric charge (\ref{Q}) also vanishes, since it is quadratic in $\chi'$. Thus, we conclude that the two main effects induced by the scalar monopole have a fully nonlinear origin.

	Now we can describe the full process developing in the range $w \in [-w_0, + w_0]$. We start in the remote past at $w= - w_0$ from a static and non-spherically symmetric geometry. While the scalar profile vanishes there, its gradient does not which ultimately generates and effective electric charge $Q$. Then, a scalar pulse is ignited which induces both a longitudinal as well as a transverse GWs which propagate to $\cI^{+}$. Along the process, this nonlinear superposition of these GWs is described by an algebraically general type I geometry.
	
	To the best of our knowledge, this non-perturbative radiative solution provides a first example of a new phenomenology emerging in the fully nonlinear regime of a higher-order scalar-tensor theory. The possibility to generate a tensorial wave from a purely scalar monopole configuration is intriguing and has to be understood as a special effect inherent to i) the higher-order terms in the Horndeski dynamics and ii) to the fully nonlinear regime. We shall come back on this last point in the end. For the moment, we would like to confirm our analysis by inspecting the different polarizations of the GWs propagating in the new solution.

	\section{Polarizations and memory effect}
	\label{D}
	
	\subsection{Polarizations from the Penrose limit}
	
	In the linear theory, where GWs are described by a small perturbation $h_{\mu\nu}$ around some background $g_{\mu\nu}$, one can read off the different polarizations of the wave from the structure of $h_{\mu\nu}$. However, for our non-perturbative solution, such an approach is no longer applicable. Instead, one can use the so-called Penrose limit to analyze the different GW polarizations present in a radiative spacetime.
	
	For a given spacetime $g_{\mu\nu}$, the Penrose limit consists in encoding the leading tidal effects experienced by a photon around its worldline $\gamma$ within a pp-wave \cite{Penrose}. The profile of the pp-wave is directly related to the leading geodesic deviation effect in the direction transverse to the geodesic motion. This procedure is remarkable in that it keeps track of the non-perturbative character of the initial metric. Starting from a non-perturbative radiative spacetime, it allows one to identify in a simple way the different polarizations propagating in this geometry by simply reading the profile off the associated pp-wave. We refer to \cite{Blau:2003dz,Blau:2006ar} for a detailed and pedagogical exposition of the Penrose limit.
	
	To proceed, one first constructs null Fermi coordinates adapted to a given null geodesic. The change of coordinates is in the form of a Taylor expansion in the spacelike hypersurface orthogonal to the $4$-velocity of the geodesic $\bar{\gamma}$. To that end, we consider the PTF $E^{\mu}_A$ introduced in the previous section and denote the affine parameter of this geodesic $W$ such that $\ell^{\mu} = E^{\mu}_{W} \rd W$.  Following the construction reviewed in \cite{Blau:2006ar}, we denote the null Fermi coordinates $X^{A} =(W, V, X^i)$ with $i\in\{1,2\}$. In terms of the initial coordinates $x^{\mu} = (w, x^a)$ with $x^a = (\rho, x,y)$ the coordinates in the hypersurface orthogonal to the geodesic motion, the change of coordinate between the initial and Fermi coordinates is given by
	\begin{align}
		\label{Fermi}
		X^{A} & = E^{A}_{a} x^a+ E^{A}_{\mu} \bar{\Gamma}^{\mu}{}_{ab}  x^a x^b + \cO((x^a)^3) \,,
	\end{align}
	where we restrict the change of coordinates to the quadratic terms in $x^a$. Here, the overbar means that the quantity is evaluated on the lightlike geodesic of reference $\bar{\gamma}$.  Performing the change of coordinates, the metric becomes
	\begin{align}
		\rd s^2 & = 2 \rd W \rd V + \delta_{AB} \rd X^A \rd X^B - \bar{R}_{W A W B} (W)X^A X^B \rd W^2 \nn \\
		& \;\;\;  - \frac{4}{3} \bar{R}_{W ABC}(W) X^A X^C\rd W \rd X^B  - \frac{1}{3} \bar{R}_{ABCD} (W)X^A X^{C} \rd X^B \rd X^D \nn \\
		\label{fermi}
		& \;\;\; + \cO(X^3) \,,
	\end{align}
	where higher and higher-order in the expansion provide additional information on the gravitational field around the geodesic $\bar{\gamma}$. This expansion can be organized as follows. Considering the conformal transformation of the coordinates together with a rescaling of the metric $g_{AB} \rightarrow \lambda^{-2} g_{AB}$, it can be shown that the Weyl scalars $\Psi_i$ scale as $\cO(\lambda^{4-i})$ for $i\in\{0,...,4\}$. The Penrose limit amounts at approximating the geometry around the null geodesic by a Petrov type N geometry, i.e. keeping only the contribution from $\Psi_4$ at leading order in the $\lambda$ expansion \cite{Blau:2006ar, Kunze:2004qd}. This selects the first two lines in \eqref{fermi} given by
	\begin{align}
		\label{pp}
		\rd s^2 & = 2\rd W \rd V  + \delta_{AB} \rd X^A \rd X^B - H_{AB} (W) X^A X^B \rd W^2 \,,
	\end{align} 
	with the wave profile given by
	\be
	\label{prof}
	H_{AB}(W) \equiv \bar{R}_{WAWB}(W) = \bar{R}_{\mu\nu\rho\sigma} E^{\mu}_{W} E^{\nu}_A E^{\rho}_{W} E^{\sigma}_B \,.
	\ee
	This metric corresponds to a pp-wave in the standard Brinkmann coordinates and the matrix $H_{AB}$ is the key quantity encoding the polarizations of the nonlinear gravitational plane wave. In full generality, the wave profile can be decomposed as
	\begin{align}
		\label{profile}
		H_{AB} X^A X^B & = H_{\circ} (X^2 + Y^2) + \left[ H_{+} (X^2 - Y^2) +  H_{\times} X Y  \right] \,,
	\end{align} 
	where we have denoted the spatial coordinates $X^A=(X,Y)$. The quantity $H_{\circ}$ encodes the trace contribution, associated with the longitudinal degree of freedom, while $H_{+, \times}$ encode the two traceless tensorial polarizations of the wave. Let us emphasize that this process keeps the nonlinear nature of the original metric (\ref{metdis}) intact, in the sense that we never impose that $H_{AB}$ be small. We refer to \cite{Blau:2006ar} for further details.
	
	Thus, at leading order, once we have identified a PTF, constructing the pp-wave geometry associated to our radiative geometry reduces to computing the projection \eqref{prof}. 
	Proceeding in that way, we find that the plane wave describing the Penrose limit of our spacetime for the chosen geodesic $\bar{\gamma}$ contains three polarizations. The two tensorial polarizations can be expressed in terms of the non-vanishing Weyl scalar $\tilde{\Psi}_0$ of the original metric as
	\begin{align}
		\label{tens}
		H_{+} = \frac{1}{2} \text{Re}(\bar{\tilde{\Psi}}_0) \,, %\frac{P \chi^4}{4(\rho^2 - \chi^2)^5} \left[P(x, y) (k_{xx} - k_{yy}) + 2 k_x P_x - 2 k_y P_y\right] \,,\\
		\qq{and}
		H_{\times} = \frac{1}{2} \text{Im}(\bar{\tilde{\Psi}}_0) \,, %\frac{P\chi^4}{2(\rho^2 - \chi^2)^5} \left[P_y k_x + P_x k_y + P k_{xy}\right]
	\end{align}
	where $\bar{\tilde{\Psi}}_0$ means that $\tilde{\Psi}_0$ is evaluated on the reference geodesic $\bar{\gamma}$. 
	
	The scalar or trace polarization $H_{\circ}$ has a complicated expression and we provide it in appendix~\ref{app:trace-polar}. The Penrose limit confirms that at the fully non-perturbative level, the new solution contains both scalar and tensorial GWs. Finally, we stress that while the scalar sector $H_{\circ}$ receives contributions from all orders in $B_0$, the tensorial modes are triggered at quadratic order in $B_0$, following the pattern found for the Weyl scalar $\tilde{\Psi}_0$ and for the shear $\tilde{\sigma}$. 
	
	Now, we turn to the analysis of the geodesic motion of test particles and the associated memory effect induced by the pulse.

	\subsection{Geodesic motion and memory effect}
	
\begin{figure}[!htb]
	\centering
		\begin{subfigure}{0.4\textwidth}
			\includegraphics{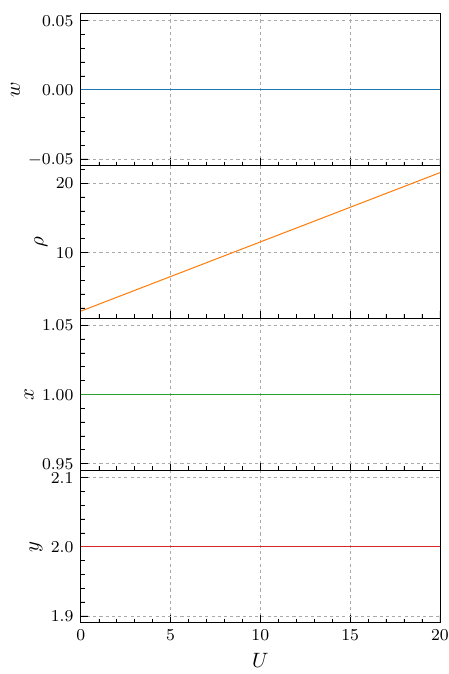}
			\caption{$B_0 = 0$ (GR)}
			\label{fig:geodesic-GR}
		\end{subfigure}
		\hspace{0.5cm}
		\begin{subfigure}{0.4\textwidth}
			\includegraphics{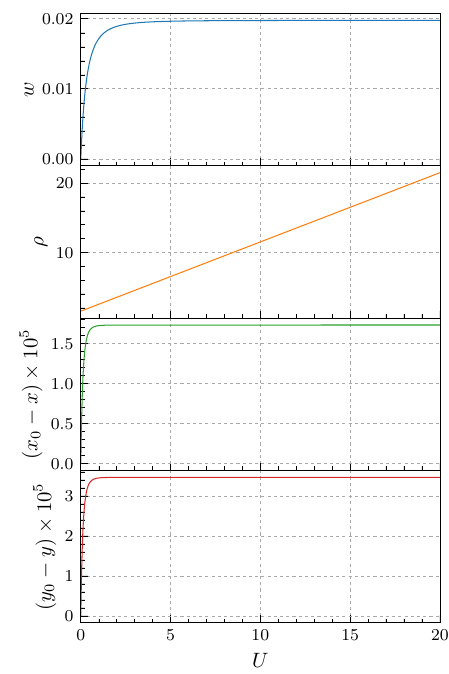}
			\caption{$B_0 = 0.1$}
			\label{fig:geodesic-dis}
		\end{subfigure}
		\caption{Evolution of the coordinates $w$, $\rho$, $x$ and $y$ along a geodesic parallel to $\ell^\mu$ for $w_0 = 0$, $\rho_0 = 1.5$, $x_0 = 1$ and $y_0 = 2$. We have set $\alpha = \alpha_0$, $C_0 = 1$.}
		\label{fig:geodesic-integration}
	 \end{figure}

	As a first step, we consider the geodesic equation in the exact solution (\ref{metdis}), i.e.
	\be
	\tilde{\ell}^{\alpha} \tilde{\nabla}_{\alpha} \tilde{\ell}^{\mu} =0 \,, \qquad \text{with} \qquad \tilde{\ell}^{\mu} = \frac{\rd x^{\mu}}{\rd U} \,.
	\ee
	%In order to visualize the quantities appearing in the matrix $H_{AB}$, one needs to evaluate their expressions on a reference null geodesic. To do this, one must therefore solve the geodesic equation. 
	Since we already have the explicit expression for the vector field parallel to the geodesic, $\tilde{\ell}^\mu$, we simply need to solve the equations
	\begin{align}
		&\dv{w}{U} = \tilde{\ell}^w \,,  &&\dv{\rho}{U} = \tilde{\ell}^\rho \,, 
		&&\dv{x}{U} = \tilde{\ell}^x \,, && \dv{y}{U} = \tilde{\ell}^y \,,
		\label{eq:geodesic-l}
	\end{align}
	with initial conditions 
	\begin{equation}
		w(0) = w_0 \,,\quad \rho(0) = \rho_0 \,,\quad x(0) = x_0 \,, \qq{and} y_0 = y_0 \,.
	\end{equation}
	This can easily be done numerically, and the results of the integration of equations~\eqref{eq:geodesic-l} are given in Fig.~\ref{fig:geodesic-integration}. One can observe that when $B_0$ is nonzero, the coordinates $w$, $x$ and $y$ are no longer constants: they evolve quickly when the lightlike particle is located at small radii and then settle down to a constant value. From a qualitative point of view, the disformed geodesic motion is modified by a small displacement on the hypersurface with the orthogonal vector $\tilde{\ell}^{\mu}$. However, it would be misleading to interpret this constant shift as a displacement memory effect as the analysis of the memories has to be done in the local inertial frame of the test particle, i.e. by constructing suitable Fermi coordinates and measuring the spatial distance in this specific frame.
	
	In order to discuss the memory effect induced by the scalar pulse, we consider instead the Penrose limit of our exact solution around this geodesic $\gamma$, i.e.
	\begin{align}
		\label{ppp}
		\rd s^2 & = 2\rd W \rd V  + \delta_{AB} \rd X^A \rd X^B - H_{AB} (W) X^A X^B \rd W^2 \,,
	\end{align} 
	This geometry (\ref{ppp}) encodes the leading tidal effects experienced by the lightlike observer with worldline $\gamma$. The coordinates (\ref{Fermi}) are the Fermi null coordinates, allowing one to analyze the leading memory effects induced by the GWs in the local inertial frame of the lightlike observer. Starting from this pp-wave geometry, we now solve the geodesic motion and compute the relative distance and relative velocity between two nearby photons. Notice that the Euclidean spatial distance between these two test particles can make sense only in this Fermi frame. We refer the reader to \cite{Zhang:2018srn, Flanagan:2019ezo} for different investigations of velocity memory effects in pp-wave geometries and \cite{Divakarla:2021xrd} for a derivation in the context of the wave form sourced by the coalescence of a binary black hole merger.
	
\begin{figure}[!htb]
		\centering
		\includegraphics{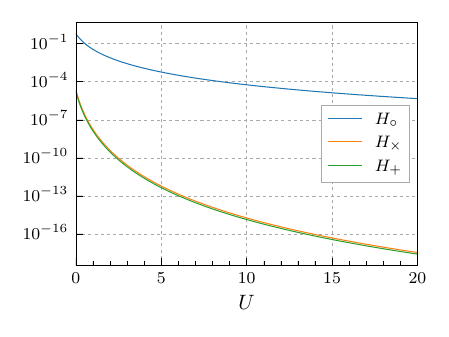}
		\caption{Plot of the amplitudes $H_\circ$, $H_\times$ and $H_+$ as functions of $U$ along the geodesic described in Fig.~\ref{fig:geodesic-integration}. One can notice that the amplitude of the tensorial waves is much smaller than that of the scalar waves.}
		\label{fig:pulse_plots}
	\end{figure}

	To proceed, we first evaluate the pulse shapes $H_\circ(W)$, $H_\times(W)$ and $H_+(W)$ seen by a photon traveling along this geodesic. This is shown in Fig.~\ref{fig:pulse_plots}. One can observe from these plots that the amplitude of the scalar wave $H_\circ$ is much bigger than the amplitudes of the tensorial waves $H_\times$ and $H_+$ generated from the scalar monopole.
	
\begin{figure}[!htb]
		\centering
		\includegraphics{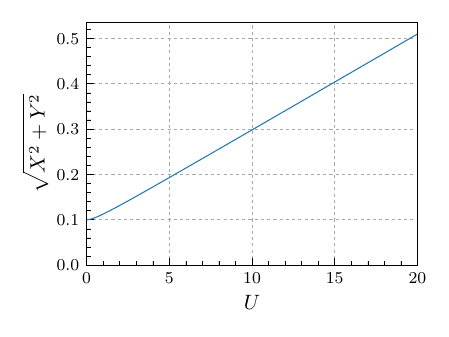}
		\caption{Plot of the distance $\sqrt{X^2 + Y^2}$ between a photon following the geodesic $\bar{\gamma}$ and a photon following the same geodesic with a small initial deviation $X = X_0$, $Y = Y_0$. The computations are done with $X_0 = Y_0 = 0.1/\sqrt{2}$.}
		\label{fig:memory-effect}
	\end{figure}

	With these forms for the components of $H_{AB}$, we now compute the distance between two photons traveling initially close to the geodesic $\gamma$. Indeed, the geodesic equations are given by
	\begin{align}
		\ddot{W} & = 0 \,, \\
		\ddot{V}  & = H_{\circ} ( \dot{X}^2 + \dot{Y}^2 ) + H_{+} ( \dot{X}^2 - \dot{Y}^2 ) + 2 H_{\times} \dot{X} \dot{Y} \,,\\
		\ddot{X} & = H_\circ X + H_+ X + H_\times Y \,,\\
		\ddot{Y} & = H_\circ Y - H_+ Y + H_\times X \,.
	\end{align}
	where a dot refer to a derivative w.r.t. the affine parameter $U$. It is direct to see that $W = c U + W_0$. Choosing without loss of generality $W_0 =0$ and $c = 1$, we identify the affine parameter with the coordinate $W$. The three remaining equations can be solved numerically. Let us now focus on the geodesic motion restricted to the two-dimensional plane $(X,Y)$. In the case of a pp-wave, the geodesic equation for $(X,Y)$ coincides with the geodesic deviation equation. Thus, by solving for the motion of $(X,Y)$, one can compute the evolution of the distance $\sqrt{X^2 + Y^2}$ between a photon following exactly $\bar{\gamma}$ (having $X = 0$ and $Y = 0$ for any $U$) and a photon starting at given initial values $X_0$ and $Y_0$ for $X$ and $Y$ (respectively). The evolution of the distance $\sqrt{X^2 + Y^2}$ is given on Fig.~\ref{fig:memory-effect}.
	
	One observes that the relative distance between the two null test particles grows linearly in the $W$-coordinate. It follows that for two photons with a vanishing initial relative velocity, the scalar pulse induces a constant shift in their relative velocity, an effect known as the velocity memory effect.  One can show that the effect of the transverse wave is much smaller than the longitudinal component, as expected. As a consequence, the correction to the velocity memory effect induced by the pure longitudinal wave is negligible. Notice that this memory effect has been computed using a truncation of the full geometry. Nevertheless, since this truncation contains the leading tidal effects around any observer describing the geometry around its worldline using the adapted Fermi coordinates, this approach captures the relevant contribution from our radiative solution which is encoded in the explicit expressions of the functions $(H_{\circ}, H_{+}, H_{\times})$.

	\section{Discussion}
	
	\label{disc}
	
	We have constructed a new exact radiative solution of a specific Horndeski theory which allows one to investigate the phenomenology resulting from the scalar-tensor mixings in the fully nonlinear regime. To the best of our knowledge, it provides the first non-perturbative solution beyond pp-waves of this kind in the context of Horndeski gravity. This construction has been achieved by using the disformal solution-generating method, applying the procedure to the seed solution (\ref{scalprof}) for the Einstein-Scalar system presented by Tahamtan and Svitec in \cite{Tahamtan:2015sra, Tahamtan:2016fur}. 
	
	At the level of the seed solution, the first result of this work consists in having numerically solved for the profile of the functions $K(x,y)$ and $P(x,y)$. The solution for $K(x,y)$ describes a deformation of the two-sphere which is localized at the equator. With this solution for $K(x,y)$ at hand, we have been able to analyze in details the nonlinear GWs propagating in this geometry. The solution describes a scalar pulse whose effect on the curvature of the two-sphere consists in a pure time-dependent rescaling. It can be interpreted as a longitudinal GW. In this solution, the scalar field is a pure time-dependent monopole controlled by the scalar charge $C_0$ (or $\alpha$). When this scalar charge vanishes, the geometry reduces to the flat Minkowski spacetime, underlying the fact that the only source of curvature is the gradient of this scalar monopole. The analysis of this radiative solution confirms the expectation: this monopole source only generates a breathing mode. This is also consistent with the fact that this GR solution belongs to the Robinson-Trautman family of Petrov type II. It follows that this geometry admits a shear-free, twist-free, and purely expanding congruence of null geodesics.
	
	Performing the disformal transformation, the Einstein-Scalar seed theory \eqref{EinS} is mapped to the shift-symmetric Horndeski action given by \eqref{action-Horndeski}. The disformal map then allows one to investigate how the scalar-tensor mixings resulting from the higher-order terms in this action, parametrized by $B_0$, respond to the pure scalar monopole. The disformal solution \eqref{metdis} thus provides an exact radiative solution of this Horndeski theory which describes the nonlinear GWs generated by the simple scalar source controlled by the scalar charge $C_0$. Remarkably, we have shown that contrary to the GR solution, the scalar monopole generates both a longitudinal and a tensorial wave. The presence of shear is a new effect whose origin can be traced back to the higher-order terms of the Horndeski action, and thus to the complicated scalar-tensor mixings in this Horndeski theory. Indeed, the expression for the shear \eqref{shear} shows that it is controlled by both the disformal parameter $B_0$ and the scalar charge $C_0$. This new nonlinear effect is the second main result of this work.
	
	It is worth emphasizing that this shear is not a gauge artifact. Indeed, the key point is that contrary to the GR solution, one can show that constructing a PTF requires fixing all the six free functions encoding the freedom in performing Lorentz transformations. Once this PTF is achieved, characterized by the spin coefficients $\tilde{\kappa} = \text{Re}(\tilde{\epsilon}) = \tilde{\pi}$, one can show that $\tilde{\sigma} \neq 0$. See \cite{letter} for a detailed derivation. Thus, contrary to the GR case, where one can construct a shear-free PTF, the disformed solution forces the PTF to exhibit a non-vanishing shear. Since any measurement of a GW detector has to be realized within such PTF, we conclude that the new shear is a physical observable effect.
	
	In order to analyze the polarization of the propagating GW in the non-perturbative manner, we have derived the leading tidal effects experienced by a photon. This has been achieved by applying the Penrose limit procedure to our non-perturbative radiative solution, deriving the associated pp-wave geometry \eqref{pp} which describes the neighborhood of the null geodesic followed by the photon. This non-perturbative approach allows one to read off the different polarizations from the wave profile \eqref{profile} of the pp-wave. We have confirmed via this procedure that a shear is indeed ignited by the scalar pulse, whose qualitative behavior agrees with the full solution. The explicit form of the polarizations $(H_{+}, H_{\times})$ was provided in \eqref{tens} and the profiles of the three polarizations $(H_{\circ}, H_{+}, H_{\times})$ are plotted in Fig.~\ref{fig:pulse_plots}. In the end, this last check confirms that the disformed solution carries a nonlinear superposition of a breathing mode (corrected w.r.t. to the GR solution) and a new tensorial mode. The explicit expressions for the different quantities have been presented in an expansion in terms of the disformal parameter $B_0$ for simplicity. Doing so, we have shown that non-vanishing shear $\tilde{\sigma}$ (and the associated polarizations $H_{+}$ and $H_{\times}$) shows up at quadratic order in $B_0$.
	
	Finally, we have solved for the geodesic motion in the full solution. As shown in Fig.~\ref{fig:geodesic-integration}, while the geodesic motion is purely radial in the GR solution, the corrections parametrized by $B_0$ modify this behavior by inducing a small and quick shift in the three other coordinates. This manifests through a small deviation of the position of the photon on the topological sphere. However, this shift cannot be interpreted as a displacement memory effect per se, since this effect can be accounted only be measuring the relative distance between two nearby test particles within the adapted Fermi coordinates. To analyze this effect, we have used instead the approximation of our exact solution obtained via the Penrose limit. Within this pp-wave, where the Brinkmann coordinates $(W,V, X,Y)$ stand as Fermi coordinates, we have computed the Euclidean distance between two nearby photons located at the equator. As expected from this pp-wave approximation, the resulting distance grows linearly with the $W$-time, demonstrating the existence of a velocity memory effect induced by the scalar pulse. This last step concludes the detailed analysis of this new exact radiative solution derived in this work.
	
	Let us now comment on future research directions. While the construction and the analysis of new exact radiative solutions provide a simple road to reveal new effects in modified gravity, such as the ones described in this work, it does not allow one to fully characterize the mechanisms responsible for this new phenomenology. Indeed, while this exact solution demonstrates the generation of shear from a pure scalar monopole source in Horndeski gravity, the effects of the higher-order terms remain hidden at the level of the solution. In order to refine our understanding of these effects, it is crucial to extend the Newman-Penrose approach to scalar-tensor theories.  In particular, it would be helpful to derive the modified equations for the optical scalars within the context of Horndeski gravity. This will be presented elsewhere.
	
	Additionally, it would be useful to investigate how the asymptotic symmetries near $\cI^{+}$ are modified in the context of higher-order scalar-tensor theories. The presence of the new terms in the dynamics will certainly modify the expressions for the charges and the fluxes entering the flux-balance laws at $\cI^{+}$. More importantly, it remains to be understood if the new type of conformal or disformal coupling can modify the falls off behavior of the fields in asymptotically flat spacetimes, which partially dictate the form of the asymptotic symmetries. These questions have already been addressed in the context of the Brans-Dicke theory in \cite{Tahura:2020vsa, Hou:2020wbo, Hou:2020tnd, Hou:2020xme, Seraj:2021qja}. Generalization to higher-order scalar-tensors theories such as DHOST theories would provide an important step to further characterize the radiative regime of these theories beyond GR at the fully nonlinear level.
	
	Finally, let us come back to the new solution presented in this work and suggest more speculative directions. Besides adding to the few known exact solutions of this kind in higher-order modified gravity, this specific solution provides a simple enough framework to witness a new effect induced by the scalar-tensor mixing in the fully nonlinear regime. If higher-order scalar tensor theories can operate at sufficiently high energy, such mechanisms might be relevant for the very early universe cosmology, prior to the inflationary phase where strong GWs should exist and have to be treated in a fully nonlinear and non-perturbative approach. While highly speculative at this stage, whether amplification of such tensorial waves can have an impact on the cosmological observables at later time in such modified gravity scenario provides an interesting question. Indeed, exact non-perturbative cosmological GWs solutions have been known for a long time in GR \cite{Belinsky:1979wi, Carr:1983jzn, Verdaguer:1986bq, Belinski:2017luc}, with interesting consequences for dark energy observables recently discussed in  \cite{Belinski:2017luc}. Whether one can construct a modified version of the solution presented in this work which can be relevant for the early universe cosmology provides an interesting future research direction.

	%===============================================================================
	\acknowledgments
	
	The work of MAG was supported by Mar\'{i}a Zambrano fellowship.
	
	%===============================================================================
	%==============================================================================
	
	\appendix

	\section{Numerical computation of $P(x,y)$}
	\label{app:resolution-P}
	
	In this appendix, we describe the numerical method used to compute a solution to Eqs~\eqref{eq:ang-K}. We change variables to express the main equation for $P$ in spherical coordinates $(\theta, \varphi)$. The change of variables is given by
	\begin{equation}
		\label{cood}
		\sqrt{x^2 + y^2} = \cot\frac{\theta}{2} \,, \qq{and} \varphi = \arg(x + iy) \,.
	\end{equation}
	Furthermore, we assume that $P$ depends only on $\theta$. The resulting equation takes the form
	\begin{align}
		&\dv[4]{f}{\theta} + 2 \cot\theta \dv[3]{f}{\theta} - \frac{1}{f} \Big(\cot^2\theta f - 2 \cot\theta \dv{f}{\theta} + \dv[2]{f}{\theta}\Big) \dv[2]{f}{\theta} \nonumber\\
		&+ \frac{\cot\theta}{f} \Big(\Big(2+ \frac{1}{\sin^2\theta}\Big) f - \cot\theta \dv{f}{\theta}\Big) \dv{f}{\theta} - \frac{\alpha^2}{f^3} = 0 \,,
		\label{eq:eq-P-spherical}
	\end{align}
	with $f = \flatfrac{P}{P_0}$, $P_0$ being defined in Eq.~\eqref{P0}.
	
	Since Eq.~\eqref{eq:eq-P-spherical} is singular at $\theta = 0$ and $\theta = \pi$, it is not possible to impose boundary conditions and simply integrate the equation. However, one can solve the equation asymptotically at both of these points. Let us take the example of $\theta = 0$. We look for solutions of the form
	\begin{equation}
		f(\theta) = 1 + \frac{a_2}{2!} \theta^2 + \frac{a_4}{4!} \theta^4 +... \,.
	\end{equation}
	This allows us to impose $f = 1$, or $P = P_0$ at the North pole. The solution is expected to be even-parity around $\theta = 0$ and that is why we did not include the odd-parity terms. After injecting this ansatz in Eq.~\eqref{eq:eq-P-spherical}, one obtains a set of nonlinear equations that can be solved for the $a_i$. This allows us to obtain the solution $f_\text{asymp}$ close to $\theta = 0$. We then set $\theta_0 = 0.01$ and solve Eq.~\eqref{eq:eq-P-spherical} from $\theta = \theta_0$ to $\theta = \flatfrac{\pi}{2}$, using as boundary conditions at $\theta = \theta_0$ the values of $f_\text{asymp}$ and its derivatives. As a check of consistency, we verify that injecting the solution in the original equation yields zero.
	
	One can apply this procedure for any value of $\alpha$. However, one would expect $f$ (and $P$) to be symmetrical when $\theta$ is changed into $\pi - \theta$. In order to impose this, we require
	\begin{equation}
		\left.\dv{f}{\theta}\right|_{\theta = \flatfrac{\pi}{2}} = 0 \,.
	\end{equation}
	This is not possible for any value of $\alpha$. For instance, one finds that in the case $\alpha^2 \simeq 1.541$, such a solution exists. Let $\alpha_0$ be the corresponding value of $\alpha$. The final solution for $f$ is then built by symmetrizing the solution between $\theta = 0$ and $\theta = \flatfrac{\pi}{2}$. It is shown on Fig.~\ref{fig:K} alongside the corresponding function $K$. One should note that this solution is not $\mathcal{C}^\infty$ but rather $\mathcal{C}^2$. As a consequence, $K$ is only of class $\mathcal{C}^0$.

	\section{Spin coefficients and Petrov classification}\label{App-D}
	
	In this appendix, we present definitions of spin coefficients and Weyl scalars. 
	To define the PTF, we need the definition of the following optical scalars \cite{Chandrasekhar:1985kt,Stephani:2003tm}
	\begin{align}\label{spin-coefficients}
	\kappa &\equiv -m^{\alpha}\ell^\mu \nabla_\mu\ell_{\alpha} \,,
	&&\epsilon \equiv \frac{1}{2}(\bar{m}^{\alpha} \ell^\mu \nabla_\mu m_{\alpha} - n^{\alpha}\ell^\mu \nabla_\mu \ell_{\alpha}) \,,
	&&\pi \equiv \bar{m}^{\alpha}\ell^\mu \nabla_\mu n_{\alpha} \,.
	\end{align}	
	The shear, expansion, and twist are defined as
	\begin{align}\label{sigma-omega-theta}
	\sigma &\equiv -m^{\alpha}m^\mu \nabla_\mu\ell_{\alpha} \,,
	&&\Theta \equiv {\rm Re}\left[ m^{\alpha}\bar{m}^\mu \nabla_\mu\ell_{\alpha} \right] \,,
	&&\omega \equiv {\rm Im}\left[ m^{\alpha}\bar{m}^\mu \nabla_\mu\ell_{\alpha} \right] \,.
	\end{align}

	The Weyl scalars are defined as~\cite{Chandrasekhar:1985kt,Stephani:2003tm}
	\begin{eqnarray}\label{WeylS0}
		{ \Psi}_0 & =& {C}_{\alpha\beta\mu\nu} {\ell}^{\alpha} { m}^{\beta} {\ell}^{\mu} { m}^{\nu} \,,\\ \label{WeylS1}
		{ \Psi}_1 & =& {C}_{\alpha\beta\mu\nu} { \ell}^{\alpha} { n}^{\beta} { \ell}^{\mu} { m}^{\nu}  \,, \\ \label{WeylS2}
		{ \Psi}_2 & =& { C}_{\alpha\beta\mu\nu} { \ell}^{\alpha} { m}^{\beta} {\bar m}^{\mu} { n}^{\nu}  \,, \\ \label{WeylS3}
		{ \Psi}_3 & =& { C}_{\alpha\beta\mu\nu} { n}^{\alpha} { \ell}^{\beta}  { n}^{\mu} {\bar m}^{\nu} \,, \\  
		{ \Psi}_4 & =& { C}_{\alpha\beta\mu\nu} { n}^{\alpha} {\bar m}^{\beta} { n}^{\mu} {\bar m}^{\nu} \,, \label{WeylS4}
	\end{eqnarray}
	where $C_{\alpha\beta\mu\nu}$ is the Weyl tensor. 
	The Weyl scalars are not Lorentz-invariant and they are defined up to the Lorentz transformations. The Petrov classification is based on the Lorentz-invariant quantities which constructed out of the Weyl scalars as follows
	\begin{align}\nonumber
		&{ I} \equiv { \Psi}_0 { \Psi}_4 - 4 { \Psi}_1 { \Psi}_3 
		+ 3 { \Psi}_2^2 \,, \hspace{.5cm}
		{ J} \equiv 
		\begin{vmatrix}
			{ \Psi}_4 & { \Psi}_3 & { \Psi}_2 \\ 
			{ \Psi}_3 & { \Psi}_2 & { \Psi}_1 \\ 
			{ \Psi}_2 & { \Psi}_1 & { \Psi}_0
		\end{vmatrix} \,, \\ \nonumber
		& { M} \equiv { \Psi}_4^2 { \Psi}_1 
		- 3 { \Psi}_4 { \Psi}_3 { \Psi}_2 + 2 { \Psi}_3^3 \,, \hspace{.5cm}
		{ L} \equiv { \Psi}_4 { \Psi}_2 - { \Psi}_3^2 \,, \\
		&{ N} \equiv 12 { L}^2 - { \Psi}_4^2 { I} 
		\,, \hspace{.5cm}
		{ D} \equiv { I}^3 - 27 { J}^2 \,.
		\label{Petrov-quantities}
	\end{align}
	From the above coordinate-independent and Lorentz-invariant quantities, one can find the Petrov type of a given geometry as shown in Table \ref{tab1}. 
	\begin{table}
		\centering
		\begin{tabular}{ |p{1cm}|p{5cm}|  }
			\hline 
			\hfil Type & \hfil Conditions \\
			\hline
			\hfil I & ${ D}\neq0$ \\
			\hfil II & ${ D} =0$, ${ I}\neq0$, ${ J}\neq0$, ${ K}\neq0$, ${ N}\neq0$ \\
			\hfil III & ${ D} =0$, ${ I} = { J} =0$, ${ K}\neq0$, ${ L}\neq0$ \\
			\hfil N & ${ D} =0$, ${ I}={ J}={ K}={ L}=0$ \\
			\hfil D & ${ D} =0$, ${ I}\neq0$, ${ J}\neq0$, ${ K}={ N}=0$\\
			\hfil O & ${ \Psi}_0={ \Psi}_1={ \Psi}_2={ \Psi}_3={ \Psi}_4=0$ \\
			\hline
		\end{tabular}
		\caption{Petrov classification.}
		\label{tab1}
	\end{table}
	
	Using the null tetrad \eqref{seednullvec1} together with the seed metric \eqref{met} in definitions \eqref{WeylS0}-\eqref{WeylS4}, we find \eqref{Psis-seed}. Substituting \eqref{Psis-seed} in \eqref{Petrov-quantities}, we find
	\begin{align}
	D=0 \,, 
	\qq{or}
	S = \frac{27 J^2}{I^3} = 1 \,,
	\end{align}
	which shows the well known result that the  Robinson-Trautman metric is algebraically special and it is type II.
	Following the similar procedure but with the disformed PTF \eqref{PTF}, it is straightforward to show that
	\begin{align}
	\tilde{D}=0 \,, 
	\qq{or}
	{\tilde S} \equiv \frac{27 {\tilde J}^2}{{\tilde I}^3} \neq 1 \,,
	\end{align}
	which shows that the disformed metric is type I. 
	Therefore, the Robinson-Trautman metric downgrades from type II to type I after performing the disformal transformation.
	
	\section{Trace polarization expression}
	\label{app:trace-polar}
	
	In this appendix, we provide the explicit expression at quadratic order in $B_0$ for the trace polarization of the pp-wave resulting from the Penrose limit:
	
	\begin{multline}
		H_\circ = \frac{\chi^2}{\left(\rho ^2-\chi^2\right)^2} - \frac{B_0}{2(\rho^2 - \chi^2)^4} \Big[4 \left(2 \rho ^2 \chi ^2+\chi ^4\right) K +\chi ' \left(2 \rho  \chi  \left(5 \rho ^2+7 \chi ^2\right)+ \log(\frac{\rho - \chi}{\rho + \chi}) \left(\rho ^2-\chi ^2\right)^2\right) \Big] \\
		\begin{aligned}[t] \qquad&+ \frac{B_0^2}{96 \chi ^2 \left(\rho ^2-\chi ^2\right)^6} \Big[3 \log(\frac{\rho - \chi}{\rho + \chi})  \left(\rho ^2-\chi ^2\right)^2 \Big(15 \rho ^4 \chi ' K +34 \rho ^2 \chi ^2 \chi ' K+47 \chi ^4 \chi ' K+2 \rho ^3 \chi  (\alpha ^2+48 (\chi ')^2) \\
			&\quad-2 \rho  \chi ^3 (\alpha ^2-48 (\chi ')^2)\Big) +2 \chi  \Big(\rho ^2 \chi ^5 \Big(768 K^2 +7 (\alpha ^2+288(\chi ')^2)\Big) -4 \chi ^7 (-48 K^2+\alpha ^2-36 (\chi ')^2) \\
			&\quad +45 \rho ^7 \chi ' K+27 \rho ^5 \chi ^2 \chi ' K+2235\rho ^3 \chi ^4 \chi ' K +1533 \rho  \chi ^6 \chi ' K+3 \rho ^6 \chi  (\alpha ^2+80 (\chi ')^2) \\
			&\quad-6 \rho ^4 \chi ^3 (\alpha ^2-240	(\chi ')^2)\Big)  +\frac32 \log(\frac{\rho - \chi}{\rho + \chi})^2 \left(\rho ^2-\chi ^2\right)^4 (\alpha ^2+16 (\chi ')^2) \Big] + \mathcal{O}(B_0^3) \,. \end{aligned} 
	\end{multline}

\providecommand{\href}[2]{#2}\begingroup\raggedright\endgroup

\end{document}